\documentclass[traditabstract]{aa}
\usepackage{graphicx}
\usepackage{color}
\usepackage[english]{babel}
\usepackage{float}
\usepackage[utf8x]{inputenc}
\usepackage{natbib}
\usepackage{txfonts}

\title{The Milky Way halo as a QSO absorption-line system}
\subtitle{New results from an \emph{HST}/STIS absorption-line catalogue \\
of Galactic high-velocity clouds}

\author{
P. Herenz \inst{1},
P. Richter\inst{1,2},
J.C. Charlton \inst{3},
\and
J.R. Masiero \inst{4}
}

\institute{
Institut f\"ur Physik und Astronomie, Universit\"at Potsdam, 
Karl-Liebknecht-Strasse 24/25, 14476 Potsdam-Golm, Germany
\email{pherenz@astro.physik.uni-potsdam.de}
\and
Leibniz-Institut f\"ur Astrophysik Potsdam (AIP),
An der Sternwarte 16, 14482 Potsdam, Germany
\and
Department of Astronomy and Astrophysics, Pennsylvania State
University, University Park, PA 16802, USA
\and
Jet Propulsion Laboratory, 4800 Oak Grove Drive, 
Pasadena, CA 91109, USA
}

\date{Accepted by A\&A on 17 Dec 2012}

\abstract{
We use archival UV absorption-line data
from \emph{HST}/STIS to statistically analyse the absorption
characteristics of the high-velocity clouds (HVCs) in the
Galactic halo towards more than 40 extragalactic background
sources. 
We determine absorption covering fractions of low- and intermediate
ions (O\,{\sc i}, C\,{\sc ii}, Si\,{\sc ii}, Mg\,{\sc ii},
Fe\,{\sc ii}, Si\,{\sc iii}, C\,{\sc iv}, and Si\,{\sc iv})
in the range $f_{\rm c}=0.20-0.70$. For detailed
analysis we concentrate on Si\,{\sc ii} absorption components
in HVCs, for which we investigate the distribution of
column densities, $b$-values, and radial velocities.
Combining information for Si\,{\sc ii} and Mg\,{\sc ii}, and using
a geometrical HVC model we
investigate the contribution of HVCs to the absorption cross section
of strong Mg\,{\sc ii} absorbers in the local Universe. We estimate 
that the Galactic HVCs would contribute on average $\sim 52$ percent
to the total strong Mg\,{\sc ii} cross section of the
Milky Way, if our Galaxy were to be observed from an
exterior vantage point. We further estimate that the mean projected 
covering fraction of strong Mg\,{\sc ii} absorption 
in the Milky Way halo and disc from an exterior vantage point is
$\langle f_{\rm c,sMgII}\rangle =0.31$ for a halo radius
of $R=61$ kpc. These numbers, together with the observed
number density of strong Mg\,{\sc ii} absorbers at low redshift,
indicate that the contribution of infalling gas clouds
(i.e., HVC analogues) in the halos of Milky Way-type galaxies 
to the cross section of strong Mg\,{\sc ii} absorbers
is $< 34$ percent.
These findings are in line with the idea that 
outflowing gas (e.g., produced by galactic winds)
in the halos of more actively star-forming galaxies
dominate the absorption-cross section of strong Mg\,{\sc ii}
absorbers in the local Universe.
}


\titlerunning{The MW halo as QSO absorber}
\begin{document}
\maketitle

%

\section{Introduction}

The so-called high-velocity clouds (HVCs) in the halo of the Milky Way
are believed to represent an important phenomenon related to the
ongoing formation and evolution of galaxies at $z=0$. HVCs (in 
their original definition) represent
neutral gas clouds seen in H\,{\sc i} 21cm emission 
that are circulating with high radial velocities 
through the inner and outer regions of the Galactic halo. HVCs in
the Milky Way and other galaxies are assumed to connect
the central regions of galaxies with the surrounding intergalactic medium (IGM).
Following the usual classification 
scheme, HVCs in the Milky Way have radial velocities of $|v_{\rm LSR}|\,>90$ km\,s$^{-1}$ 
compared to the local standard of rest (LSR). Such high velocities are 
not in agreement with a standard Galactic disc rotation model. In the velocity range 
$40$ km\,s$^{-1}<\,|v_{\rm LSR}|\,<90$ km\,s$^{-1}$ there
are the so-called intermediate velocity clouds (IVCs), which complete 
the classification of Galactic halo clouds.
Since their discovery more than 40 years ago (\cite{1966Muller}),
a lot of progress has been made in understanding the distribution, origin,
and physical properties of HVCs and IVCs (see the reviews by \cite{2006Richter} and
\cite{1998Wakker}).
QSO absorption-line measurements have
shown that HVCs span a relatively large range in metallicities from
$\sim 0.1$ to $1.0$ solar (e.g., \cite{1999Wakker}; \cite{1999Richter};
\cite{2001Richter}; \cite{2001Wakker}; \cite{2001Gibson}; \cite{2003Tripp}; 
\cite{2003Collins}; \cite{2005Richter}; \cite{2009Richter}; 
\cite{2011Shull}), indicating that HVCs and
IVCs have various origins. Some HVCs, in particular the 
``Magellanic Stream'' (MS), most likely originate from gas of
smaller satellite galaxies that are being accreted by the Milky Way.
Other HVCs possibly represent metal-deficient gas that is infalling
from the intergalactic medium. The most likely origin for the IVCs 
(which predominantly have nearly solar metallicities) is the 
``galactic fountain'' (\cite{1976Shapiro}; \cite{1990Houck}).
In the galactic fountain model hot gas is ejected
out of the Galactic disc by supernova explosions. The gas then 
cools and (partly) falls back towards the disc in the form of 
condensed, neutral gas clouds.  

Recent distance estimates of several IVCs and HVCs indicate that most of 
the IVCs appear to be located within 2 kpc from the Galactic disc,
in accordance with the scenario that IVCs represent gas structures
related to the galactic fountain (\cite{2008Wakker}; \cite{2011Smoker}). 
Most of the HVCs appear to be located at distances $<20$ kpc (\cite{2007Wakker};
\cite{2006Thom}, \cite{2008Thom}), with the prominent
exception of the MS, which most likely is located as far as 50 kpc
(\cite{1996Gardiner}). These distances indicate that HVCs
do not represent Local Group (LG) objects that are related to 
the missing dark-matter (DM) halos in the LG (\cite{1999Blitz}),
but rather indicate gas circulation processes in the immediate
environment ($d<100$ kpc) of the Milky Way. Yet, with a total H\,{\sc i} mass of
$\sim 10^8$ M$_{\odot}$, HVCs contribute 
$\sim0.7\,M_{\odot}$ yr$^{-1}$ to the Milky Way's gas-accretion rate 
(\cite{2012Richter};\cite{2004Wakker}). Clearly, detailed studies
of IVCs and HVCs are of fundamental importance for our understanding
of the past and present evolution of our Galaxy.

H\,{\sc i} 21cm observations of nearby spirals (e.g., M31, NGC\,891) 
indicate that the IVC/HVC phenomenon is not restricted to the Milky Way, but reflects
gas-circulation processes at large scales that are characteristic of
low-redshift galaxies in general (\cite{2004Thilker}; \cite{2007Oosterloo}; 
\cite{2007Fraternali}). 
However, because of the limited sensitivity and beam size of 21cm observations
of more distant galaxies
it is currently impossible to spatially resolve individual gas clouds
in the halos of galaxies beyond the Local Group.
As an alternative method, QSO absorption spectroscopy has turned
out to be a powerful technique to trace neutral and ionized gas
in the extended halos of galaxies at low {\it and} at high redshift.
For this, absorption lines from low and high ionization metals, such as 
Mg\,{\sc ii} and C\,{\sc iv} in intervening absorbers, have been
analysed extensively (\cite{1991Bergeron}; \cite{1998Charlton};
\cite{2002Steidel}; \cite{2005Nestor}; \cite{2010Kacprzak}).
However, since most of the ion transitions of interest are
located in the ultraviolet (UV), QSO absorption spectroscopy
of gaseous structures in and around galaxies and in the
intergalactic medium (IGM) at $z=0$ requires spectroscopic
data from space-based UV observatories, such as the 
{\it Hubble Space Telescope} (\emph{HST}) and the 
{\it Far Ultraviolet Spectroscopic Explorer} (\emph{FUSE}). As a
consequence, the amount and the quality (in terms of
signal-to-noise, S/N) of absorption-line data of intervening
metal absorbers rising in the halos of low-redshift
galaxies is relatively limited.

Generally, QSO absorption-line measurements indicate that the circumgalactic
environment of galaxies is characterized by a complex spatial distribution
of multiphase gas that reflects both, the gas accretion processes of galaxies
from the intergalactic medium and from merger events and the 
outflow of gaseous material from galactic winds (e.g., \cite{2007Fangano};
\cite{2012Bouche}). Yet, the exact morphological relation between 
intervening metal absorbers and the Galactic HVC population has not really 
been established.

In this paper, we reanalyse HVC absorption lines from archival
UV spectral data obtained with the {\it Space Telescope Imaging
Spectrograph} (STIS) on \emph{HST} and compare the absorption characteristics
(absorption cross section, column-density distribution function)
of several ions with that of intervening metal absorption systems 
at low redshift. We focus on the absorption 
properties of Si\,{\sc ii} and Mg\,{\sc ii} in HVCs and QSO
absorbers, as these two ions have very similar ionization
potentials and thus are particularly well suited for such
a comparison. From this comparison
we derive information on the relation between Galactic HVCs and
QSO absorbers and provide an estimate for the cross section of neutral
and ionized gas structures in the halos of galaxies (see also
\cite{2012Richter}; \cite{2011Richter2}).

Many of the HVC sightlines studied in this paper have been analysed 
in great detail by various different groups (e.g., \cite{1999Wakker};
\cite{2001Sembach}; \cite{2001Richter}; \cite{2001Gibson};
\cite{2003Tripp}; \cite{2003Collins}; \cite{2004Fox}; 
\cite{2009Richter}; \cite{2009Collins}; \cite{2009Shull})
using \emph{HST}/STIS data. These groups used 
various different spectral analysis techniques (apparent optical 
depth method, AOD; curve-of-growth; profile fitting) to derive
column densities and Doppler-parameters from their data.
In addition, some of these studies (but not all) have incorporated 
supplementary spectral data from lower-resolution instruments 
(e.g., from FUSE) and from 21cm studies of neutral hydrogen,
but the strategies of how to use these supplementary data to 
derive column densities and other parameters are quite different
among the above listed studies. As a result,
there currently does not exist an homogeneous and coherent 
sample of HVC absorption-line parameters (column densities, Doppler
parameters) for important ions obtained by multi-component 
Voigt profile fitting based on the same fitting criteria for all 
HVC sightlines. A coherent absorption-line sample is desired, however,
to provide a meaningful comparison between Galactic HVC absorbers 
and intervening QSO absorbers. This motivated us to 
reanalyse \emph{HST}/STIS data of Galactic HVCs 
using the profile-fitting technique for all HVC absorption features
detected.

%

\begin{figure*}[ht]
\centering
\resizebox{0.7\hsize}{!}{\includegraphics{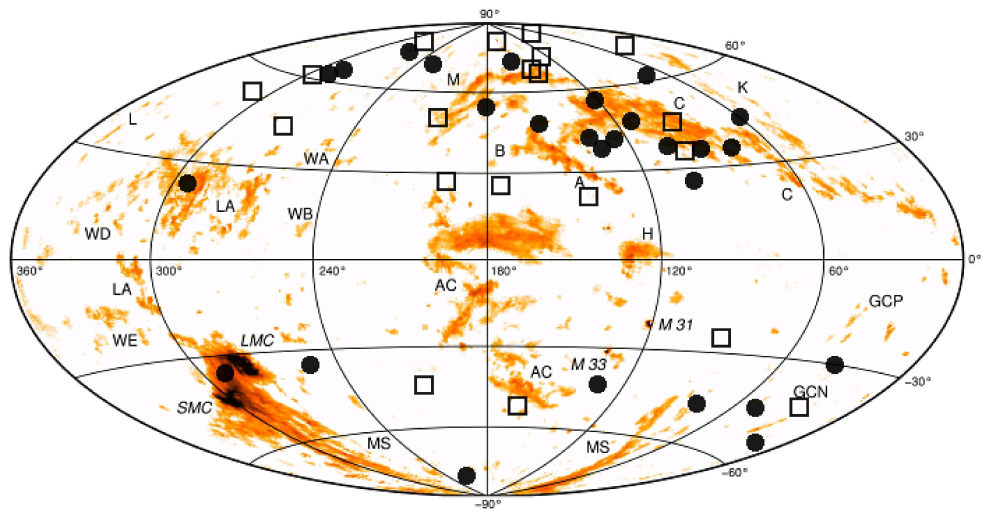}}
\caption[]{H\,{\sc i} 21cm sky map of HVC complexes, 
based on the LAB survey (\cite{2003Kalberla}, \cite{2005Kalberla}). 
The symbols mark the STIS sightlines inspected in this paper. 
Filled circles indicate sightlines with HVC detections,
whereas empty squares indicate sightlines where no HVC 
absorption was found. The LAB all-sky map was kindly provided by T.Westmeier.}
\end{figure*}

%

\begin{table}[t!]
\caption[]{Prominent ions in the \emph{HST}/STIS spectra}
\label{Ions}
\begin{tabular}{lcl}
\hline
Grating & $R=\lambda/\Delta\lambda$ & Ion \\
\hline
E140M & 45,800 & Si\,{\sc ii}, Si\,{\sc iii}, Si\,{\sc iv}, 
                 O\,{\sc i}, C\,{\sc ii}, \\
      &        & C\,{\sc iv}, Al\,{\sc ii} \\
E230M & 30,000 & Fe\,{\sc ii}, Mg\,{\sc ii} \\
\hline
\end{tabular}
\end{table}

%

\section{Data aquisition and analysis method}

Our \emph{HST}/STIS data set contains 47 sightlines through the Galactic halo 
towards QSOs and other AGNs. All spectra are publicly available in the 
Mikulski Archive for Space Telescope (MAST). Fig.\,1 shows the sky
distribution of the 47 sightlines plotted on a H\,{\sc i} 21cm
map of the Galactic HVCs from the Leiden/Argentine/Bonn
(LAB) all-sky survey (\cite{2006Kalberla}).
From an inspection of the original observing proposals we conclude that
more than 70 percent of the 47 selected AGN were selected because of their 
ultraviolet (UV) brightness and/or because of the known presence of 
intervening metal-absorption systems, as indicated by previous UV data 
from earlier \emph{HST} spectrographs with lower spectral resolution. 
We therefore can assume that the 47 sightlines are not biased towards
particular HVCs or HVC regions. Fig.\,1 indicates, however, that 
there is a clear overabundance of QSO sightlines in the northern
sky at $b>30$ deg
with both detections and non-detections of high-velocity halo
gas in the STIS data. This non-uniform sky distribution 
of the QSOs in our sample is further discussed in Sect.\,3.

All STIS spectra considered in this paper were recorded using the
E140M ($\lambda=1150-1700$ \AA) and the E230M ($\lambda=1600-3100$ 
\AA) high-resolution Echelle gratings of STIS. These
instruments provide a spectral resolution of $R\sim 45,800$ (E140M) 
and $R\sim 30,000$ (E230M), corresponding to a velocity 
resolution of $\sim7$ km\,s$^{-1}$ and $\sim10$ km\,s$^{-1}$
FWHM, respectively. 
The STIS data were reduced using the standard STIS
reduction pipeline (\cite{2002Brown}). Separate
exposures were combined following
the procedures described by \cite{2005Narayanan}.
Table \ref{Ions} lists prominent ions in the two wavelength
ranges that can be used to study the absorption properties of 
Galactic HVCs. The ion transitions considered in this study
include C\,{\sc ii} $\lambda 1334.5$,
C\,{\sc iv} $\lambda\lambda 1548.2,1550.8$,
O\,{\sc i} $\lambda 1302.2$,
Si\,{\sc ii} $\lambda\lambda 1190.4, 1193.3, 1260.4, 1304.4, 1526.7$, 
Si\,{\sc iii} $\lambda 1206.5$,
Si\,{\sc iv} $\lambda\lambda 1393.8,1402.8$,
Mg\,{\sc ii} $\lambda 2796.4,2803.5$, Fe\,{\sc ii} $\lambda\lambda 2382.8,2600.2$,
and Al\,{\sc ii} $\lambda 1670.8$. Laboratory wavelengths and 
oscillator strengths have been taken from the compilation
of \cite{2003Morton}. 

For our study we consider only HVCs, i.e., absorption features that
have radial velocities $|v_{\rm LSR}|\geq 90$ km\,s$^{-1}$,
but not the IVCs. All in all, 28 out of the 47 sightlines observed
with STIS exhibit significant HVC absorption 
features in at least one of the ions listed above. Table 2 
provides a summary of the HVC detections in our STIS
QSO sample. 
This overall detection rate is affected by the strongly
varying S/N ratios in the spectra and the 
resulting differing detection limits for the individual
ion lines (see also Sect.\,3). 

The HVC spectral features have been fitted by multi-component
Voigt profiles, from which we obtain column densities and 
Doppler-parameters ($b$-values) for the individual HVC
absorption components.
For the fitting process we have used the {\tt fitlyman} routine
implemented in the ESO-MIDAS software package (\cite{1995Fontana}).
After an initial inspection of the velocity structure in each
HVC we have simultaneously fitted all low ions (neutral species 
and singly-ionized species; e.g., O\,{\sc i}, Si\,{\sc ii}) 
in each velocity component with a single $b$-value. This 
approach is justified, as the low ions are expected to reside
in the same gas phase in HVCs where $b$ is dominated by turbulence
(i.e., the thermal contribution to $b$ is expected to be 
negligible). The absorption components of intermediate ions 
(e.g., Si\,{\sc iii}) and high ions (e.g., Si\,{\sc iv}, C\,{\sc iv}) 
were fitted independently of the low ions (leading to
other $b$-values for these ions), as these ions most likely trace
a gas phase different from that traced by the low ions.

Because many spectra have relatively low S/N,
not all HVC velocity components can be resolved with the
current \emph{HST} data set. 
Moreover, high-resolution, high-S/N optical 
spectra of HVC sightlines indicate that there often are a large
number of velocity sub-components in HVCs whose identification
would require
a spectral resolution much higher than currently provided
by space-based UV spectrographs (e.g.\cite{1999Welty}).
This systematic uncertainty is, however, not restricted to HVCs 
but is relevant also 
for the analysis of intervening metal absorbers at low $z$
using UV data with limited S/N and spectral resolution
(e.g., \cite{2004Richter2}; \cite{2011Ribaudo2}).
For the Voigt profile fitting presented in this paper
our strategy was to find the {\it minimum} number of velocity components 
(Voigt components) that are required to obtain a satisfying 
fit to the STIS HVC absoption profiles. This allows us to
compare our results to studies of intervening absorbers, for
which similar fitting strategies were chosen (e.g. \cite{2003Churchill}).

All HVC fitting results are listed in Tables B.1$-$B.5 of the Appendix.

%

\begin{table*}[ht!]
\tiny
\caption[]{Summary of QSO Sightlines and HVC Detections}
\label{S/N}
\vspace{0.3cm}
\resizebox{1.0\hsize}{!}{
\begin{tabular}{*{9}{l}}
\hline
\hline
QSO Name     & $z_{em}$& $l$ & $b$ & HVC status & HVC velocity range
             & HVC name & Relevant STIS grating & Detected ions \\
             & & (deg)   & (deg) & (yes/no) & (km\,s$^{-1}$) & (if known) & & \\
\hline
PKS\,2155$-$304 & 0.117 & 18 & $-$52 & yes & $-$138...$-$190 & ... & E140M
                & Si\,{\sc ii}, C\,{\sc ii}, Si\,{\sc iii}, Si\,{\sc iv}, C\,{\sc iv} \\
NGC\,5548       & 0.020 & 32 & $+$71 & no  & -               & -   & E140M, E230M
                & - \\
B\,2121$-$1757  & 0.110 & 33 & $-$42 & no  & -               & -   & E140M
                & - \\
Mrk\,509        & 0.034 & 36 & $-$30 & yes & $-$295...$-$310 & GCN & E140M, E230M
                & C\,{\sc ii}, Si\,{\sc iii}, Si\,{\sc iv}, C\,{\sc iv} \\
CSO\,873        & 1.010 & 38 & $+$84 & no  & -               & -   & E230M
                & - \\
PHL\,1811       & 0.192 & 47 & $-$45 & yes & $-$130...$-$270 & GCN & E140M
                & Si\,{\sc ii}, C\,{\sc ii}, O\,{\sc i}, Al\,{\sc ii},
                  Si\,{\sc iii}, Si\,{\sc iv}, C\,{\sc iv} \\
PG\,1630+377    & 1.480 & 60 & $+$43 & yes  & $-$160...$-$60        & -   & E230M
                & Mg\,{\sc ii}, Fe\,{\sc ii} \\
PG\,1444+407    & 0.270 & 70 & $+$63 & yes  & $-80$...$-90$  & - & E140M
                & Si\,{\sc ii}, C\,{\sc ii}, O\,{\sc i}, C\,{\sc iv} \\
PG\,1718+481    & 1.083 & 74 & $+$35 & yes & $-100$...$-215$ & C Extension
                & E230M & Fe\,{\sc ii} \\
NGC\,7469       & 0.016 & 83 & $-$45 & yes & $-$190...$-$400 & MS & E140M
                & Si\,{\sc ii}, C\,{\sc ii}, O\,{\sc i}, Si\,{\sc iii},
                  Si\,{\sc iv}, C\,{\sc iv} \\
3c351           & 0.372 & 90 & $+$36 & yes & $-$130...$-$230 & Complex C & E140M
                & Si\,{\sc ii}, C\,{\sc ii}, O\,{\sc i}, Al\,{\sc ii},
                  Si\,{\sc iii}, Si\,{\sc iv}, C\,{\sc iv}\\
Mrk\,290        & 0.030 & 91 & $+$48 & yes  & $-$128         & Complex C & E230M
                & Fe\,{\sc ii}\\
Akn\,564        & 0.030 & 92 & $-$25 & no  & -               & -
                & E140M & - \\
H\,1821+643     & 0.297 & 94 & $+$27 & yes & $-$130...$-$170 & Outer Arm
                & E140M & Si\,{\sc ii}, C\,{\sc ii}, O\,{\sc i},
                  Al\,{\sc ii}, Si\,{\sc iii}, Si\,{\sc iv}, C\,{\sc iv} \\
HS\,1700+6416   & 2.740 & 94 & $+$36 & no  & -               & -
                & E140M & - \\
PG\, 1634+706   & 1.337 & 103 & $+$37 & yes & $-$100...$-$215 & Complex C
                & E230M & Fe\,{\sc ii}\\
Mrk\,279        & 0.031 & 115 & $+$47 & yes & $-$150...$-$200 & Complex C south
                & E140M & Si\,{\sc ii}, C\,{\sc ii}, O\,{\sc i},
                  Al\,{\sc ii}, Si\,{\sc iii}, Si\,{\sc iv} \\
PG\,1259+593    & 0.472 & 121 & $+$58 & yes & $-$120...$-$145 & Complex C III
                & E140M & Si\,{\sc ii}, C\,{\sc ii}, O\,{\sc i},
                  Al\,{\sc ii}, Si\,{\sc iii}, C\,{\sc iv}\\
PG\,1248+401    & 1.030 & 123 & $+$77 & no  & - & - & E230M
                & - \\
Mrk\,205        & 0.071 & 125 & $+$42 & yes & $-$110...$-$230 & Complex C south
                & E140M & Si\,{\sc ii}, C\,{\sc ii}, O\,{\sc i}, Al\,{\sc ii} \\
3c249.1         & 0.310 & 130 & $+$39 & yes  & $-135$         & - & E140M
                & Si\,{\sc ii}, C\,{\sc ii}, Si\,{\sc iii} \\
PG\,0117+21     & 1.500 & 132 & $-$41 & yes  & $-$134              & - & E230M
                & Mg\,{\sc ii}, Fe\,{\sc ii}\\
NGC\,3516       & 0.009 & 133 & $+$42 & yes & $-$160...$-$170 & -
                & E140M, E230M
                & Si\,{\sc ii}, C\,{\sc ii}, Fe\,{\sc ii}, 
                  Mg\,{\sc ii}, Si\,{\sc iii} \\
PG\,1206+459    & 1.160 & 145 & $+$70  & no  & - & -
                & E230M & - \\
HS\,0624+6907   & 0.370 & 146 & $+$23  & no  & - & -
                & E140M & -\\
NGC\,4051       & 0.002 & 149 & $+$70  & no  & - & - & E140M
                & - \\
NGC\,4151       & 0.003 & 155 & $+$75  & yes & $+$120...$+$145 &...
                & E140M & Si\,{\sc ii}, C\,{\sc ii},
                  Si\,{\sc iii}, C\,{\sc iv}, Fe\,{\sc ii}, Mg\,{\sc ii} \\
Mrk\,132        & 1.760 & 159 & $+$49  & yes  & $-$140...$+$80 & -
                & E230M & Mg\,{\sc ii}, Fe\,{\sc ii} \\
NGC\,4395       & 0.001 & 162 & $+$82  & no  & - & - & E140M, E 230M
                & - \\
PKS\,0232-04    & 1.440 & 174 & $-$56  & no  & - & - & E230M
                & - \\
HS\,0747+4259   & 1.900 & 177 & $+$29  & no  & - & - & E230M
                & - \\
PG 0953+415     & 0.239 & 180 & $+$52  & yes & $-150$& Complex M
                & E140M & Si\,{\sc ii}, C\,{\sc ii},
                  Si\,{\sc iii}, Al\,{\sc ii} \\
HS\,0810+2554   & 1.510 & 197 & $+$29  & no  & -       & -
                & E230M & - \\
Ton\,28         & 0.330 & 200 & $+$53  & no  & - & -
                & E140M & - \\
PKS\,0405$-$123 & 0.570 & 205 & $-$42  & no  & - & -
                & E140M & - \\
PG\,1116+215    & 0.177 & 223 & $+$68 & yes & $+$180...$+$190 & -
                & E140M, E230M & Si\,{\sc ii}, C\,{\sc ii}, O\,{\sc i},
                  Fe\,{\sc ii}, Mg\,{\sc ii}, Si\,{\sc iii}, Si\,{\sc iv}, C\,{\sc iv} \\
Ton\,S210       & 0.117 & 225 & $-$83 & yes & $-$150...$-$235 & CHVC 224.0$-$83.4$-$197& E140M, E230M 
                & Si\,{\sc ii}, C\,{\sc ii}, O\,{\sc i}, Si\,{\sc iii},
                  Si\,{\sc iv}, C\,{\sc iv} \\
HE\,0515$-$4414 & 1.713 & 250 & $-$35 & yes & $+$120...$+$230 &- & E230M
                & Mg\,{\sc ii}, Fe\,{\sc ii} \\
PG\,1211+143    & 0.081 & 268 & $+$74 & yes & $+$169...$+$184 & ... & E140M
                & Si\,{\sc ii}, C\,{\sc ii}, O\,{\sc i}, Si\,{\sc iii}, C\,{\sc iv} \\
PKS\,1127$-$145 & 1.187 & 275 & $+$44 & no  & -               & -
                & E230M & - \\
PG\,1216+069    & 0.330 & 281 & $+$68& yes & $+210$...$+270$ & ... 
                & E140M & Si\,{\sc ii}, Si\,{\sc iii}, C\,{\sc iv} \\
NGC\,3783       & 0.010 & 287 & $+$23& yes & $+$180...$+$250 & Leading Arm (MS) & E140M, E230M
                & Si\,{\sc ii}, C\,{\sc ii}, O\,{\sc i}, Al\,{\sc ii}, Fe\,{\sc ii},
                  Mg\,{\sc ii}, Si\,{\sc iii}\\
3c273           & 0.160 & 290 & $+$64 & no  & -              & EPn   & E140M
                & - \\
RXJ\,1230.8+0115& 0.117 & 291 & $+$63 & yes & $-$216...$-$310 & ...  & E140M
                & Si\,{\sc ii}, C\,{\sc ii}, O\,{\sc i}, Si\,{\sc iii}, Si\,{\sc iv} \\
PKS\,0312$-$770 & 0.223 & 293 & $-$38 & yes & $+$160...$+$240 & MB   & E140M, E230M
                & Si\,{\sc ii}, C\,{\sc ii}, O\,{\sc i}, Fe\,{\sc ii},
                  Mg\,{\sc ii}, Si\,{\sc iii} \\
PG\,1241+176    & 1.280 & 293 & $+$80 & no  & -               & -    & E230M & - \\
PKS\,1302$-$102 & 0.290 & 309 & $+$52 & no  & -               & ...  & E140M & - \\
\hline
\end{tabular}}
\end{table*}

%

\section{Results}

\subsection{Covering fractions of individual ions}

The physical conditions in the gas, in particular the ionization conditions, 
are known to vary substantially among the Galactic HVC population.
HVCs span a large range in temperatures and gas
densities, they are subject to thermal instabilities, ram-pressure
stripping, photoionization from the Galactic disc and from the
extragalactic UV background, collisional ionization from 
hot material ejected by supernova explosions in the disc,
gas mixing processes, and other related phenomena. As a result, the 
characteristic absorption patterns of HVCs (i.e, the observed 
absorption frequency of metal ions and their relative strengths) 
can be used to constrain the physical conditions in HVCs.

All in all, we fit 67 individual high-velocity 
absorption components (Voigt components; see above) 
in our data set. Of these, 
47 components have velocity separations $\Delta v >30$ km\,s$^{-1}$
from neighbouring HVC absorption components and 
thus can be regarded as individual entities, hereafter
refered to as absorption {\it systems}.

An important parameter that characterizes the distribution of neutral
and ionized HVC gas in the Milky Way halo is the covering 
fraction, $f_{\rm c}(X)$, for each ion $X$ that is considered.
As indicated in Table 3, we define the covering fraction as
the number of sightlines that exhibit significant HVC absorption
in the ion $X$ divided by the total number of sightlines along which HVC
absorption above the limiting column density threshold 
(log $N_{\rm min}$) could be detected. The column density
threshold for each ion was calculated from the relevant 
ion transitions in the STIS wavelength range (see Sect.\,2) 
together with the local S/N ratio.

Table 3 shows the values of $f_{\rm c}$ for the ions  
O\,{\sc i}, C\,{\sc ii}, Si\,{\sc ii}, Mg\,{\sc ii}, Fe\,{\sc ii},
Si\,{\sc iii}, Si\,{\sc iv}, and C\,{\sc iv}, 
together with the limiting column densities, log $N_{\rm min}$, as
determined from our line-fitting analysis. In this way we 
obtain covering fractions for the above-listed ions between 
$0.20$ (Si\,{\sc iv}) and $0.70$ (C\,{\sc ii}, Si\,{\sc iii}). 
Assuming that the sightlines and the HVCs are randomly distributed over 
the sky with $f_{\rm c}\leq 1$, and considering
Poisson-like statistics, the total sky covering
fraction of HVC absorption in our data is 
$f_{\rm c}=0.70\pm0.15$. 
This HVC covering fraction
is in excellent agreement with the value of 
$f_{\rm c}=0.68\pm0.04$ derived by \cite{2012Lehner},
based on a much larger combined COS/STIS data set. The
good agreement with the Lehner et al.\,results indicates
that the non-uniform sky distribution of the QSOs (as 
mentioned in Sect.\,2) has no significant influence on the
determination of the HVC covering fraction from our 
STIS QSO sample.

Note that we do not consider absorption
by Al\,{\sc ii} in our statistical analysis, 
because for the Al\,{\sc ii} $\lambda 1670.8$ 
line (the only detectable Al\,{\sc ii} line in our data) there
is a gap between $+80$ km\,s$^{-1}\leq v_{\rm LSR} \leq 
+200$ km\,s$^{-1}$ at the red end of the Echelle grating.
To compare the covering fractions of the individual ions with each
other, and to relate them to H\,{\sc i} sky-covering fractions determined
from 21cm all-sky surveys, one needs to consider the relative
abundances of the elements (C, O, Si, Mg, and Fe) in HVCs.  The
ionization conditions and dust-depletion properties of the
absorbing gas can also affect the interpretation of the covering
fractions.  These factors will be considered in the subsequent
sections.

%

\begin{table}[ht]
\caption[]{Covering fractions of individual ions in HVCs}
\label{fc}
\vspace{0.5cm}
\begin{tabular}{lcccc} 
\hline
\hline
Ion          & ${\cal N}/{\cal N}_{\rm tot}\,^{\rm a}$ & $f_{\rm c}\,^{\rm b}$ 
             & log\,$N_{\rm min}\,^{\rm c}$\\
\hline
C\,{\sc ii}  & $21/30$         & $0.70$       & 13.20 \\
C\,{\sc iv}  & $12/30$         & $0.40$       & 13.00 \\
O\,{\sc i}   & $14/29$         & $0.48$       & 13.65 \\
Si\,{\sc ii} & $20/30$         & $0.67$       & 12.25 \\
Si\,{\sc iii}& $21/30$         & $0.70$       & 12.15 \\
Si\,{\sc iv} &  $6/30$         & $0.20$       & 12.90 \\
Mg\,{\sc ii} & $10/19$         & $0.53$       & 12.70 \\
Fe\,{\sc ii} & $10/21$         & $0.48$       & 12.90 \\
\hline
\hline
\end{tabular}
\vspace{0.3cm}
\noindent
\\
$^{\rm a}\,$Number of HVC detections above column-density 
threshold/total number of sightlines;\\ 
$^{\rm b}\,$covering fraction;\\ 
$^{\rm c}\,$minimum column density threshold considered.
\end{table}

%

\subsection{Si\,{\sc ii} absorption}

In our statistical analysis, we focus on Si\,{\sc ii} absorption in HVCs.
The STIS E140M data contain five Si\,{\sc ii} transitions
(at $\lambda1190.4$, $\lambda1193.3$, $\lambda1260.4$, 
$\lambda1304.4$, and $\lambda1526.7$) that span a large range
in oscillator strengths ($f=0.133$ for Si\,{\sc ii} 
$\lambda1526.7$ and $f=1.176$ for Si\,{\sc ii} $\lambda1260.4$; 
\cite{2003Morton}). Our simultaneous
fitting of these lines therefore provides particularly reliable values 
for $N$(Si\,{\sc ii}) and $b$(Si\,{\sc ii}) in both strong
and weak HVC absorption components. The ionization
potential of Si\,{\sc ii} ($E_{\rm Si\,II}=16.4$ eV) is very 
similar to that of Mg\,{\sc ii} ($E_{\rm Mg\,II}=15.0$ eV),
suggesting that both ions trace the same gas phase in HVCs.
In addition, the cosmic abundances of Si and Mg are almost
identical (log\,(Si/H$)_{\odot}=-4.44$ and 
log\,(Mg/H$)_{\odot}=-4.42$, assuming solar relative 
abundances from \cite{2005Asplund}).
Because Mg\,{\sc ii} is the most commonly used ion to study
circumgalactic gas at low and intermediate redshift in optical 
quasar spectra (e.g., \cite{2008Kacprzak}; \cite{2012Bouche}), 
the absorption properties of Si\,{\sc ii} and Mg\,{\sc ii} in HVCs can be 
directly compared to the statisitical properties of intervening 
Mg\,{\sc ii} absorbers at low redshift (see Sect.\,4.4).

The covering fraction for HVC Si\,{\sc ii} absorption in the halo
is $f_{\rm c}$(Si\,{\sc ii}$)=0.67$ for 
log $N$(Si\,{\sc ii}$)\geq 12.25$ (see previous section; Table 3).
For comparison, the filling factor of H\,{\sc i} in HVCs
derived from 21cm surveys is $f_{\rm c}$(H\,{\sc i}$)\approx 0.15$
for log $N$(H\,{\sc i}$)\geq 18.3$ and $f_{\rm c}$(H\,{\sc i}$)\approx 0.30$
for log $N$(H\,{\sc i}$)\geq 17.8$ (\cite{2004Wakker}.
The higher detection rate of Si\,{\sc ii} absorption 
compared to H\,{\sc i} emission suggests that
more than half ($0.37/0.67=0.55$) of the high-velocity Si\,{\sc ii} absorbers 
trace neutral and ionized gas in the halo below the
typical detection limit of current 21cm observations 
at log $N$(H\,{\sc i}$)< 17.8$.

HVCs that are detected
in metal absorption without having an H\,{\sc i} 21cm
counterpart commonly are referred to as ``low-column
density HVCs'' (LCDHVCs; \cite{2009Richter}).

%

\begin{figure}
\centering
\resizebox{1.0\hsize}{!}{\includegraphics[angle=-90]{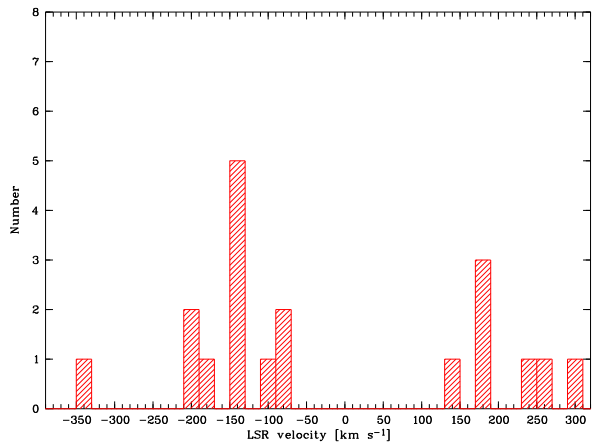}}
\caption[]{Distribution of LSR velocities of all detected 
Si\,{\sc ii} HVC absorption systems.}
\label{vLSR}
\end{figure}

%

\subsubsection{Radial velocities}

In Fig.\,2 we show the distribution of LSR velocities of 
the 19 HVCs for which Si\,{\sc ii} absorption was 
detected and accurately measured (only high-velocity 
Si\,{\sc ii} absorption towards NGC\,3516 is not considered 
here because of the low data quality). Absolute
values for $v_{\rm LSR}$ range between 
$|v_{\rm LSR}|=90$ and $370$ km\,s$^{-1}$. The highest
velocity absorber is found towards NGC\,7469 and
is related to the Magellanic Stream (Table 2).
Note that the region between $v_{\rm LSR}=-90$ to 
$+90$ km\,s$^{-1}$ (i.e., the IVC velocity regime) 
is not considered in this study.

Out of these 19 sightlines, 12
(63 percent) show absorption at negative velocities. 
One may argue that large HVC complexes at negative
velocities (such as Complex C) together with the 
limited sample size leads to an observational bias 
towards negative velocities. However, optical observations
of Ca\,{\sc ii} absorption in HVCs, based on a
ten-times larger data sample of randomly distributed
QSO sightlines, also indicate
that the majority of the neutral HVC absorbers 
exhibit negative radial velocities (\cite{2008BenBekhti}, \cite{2012BenBekhti}). 
This velocity distribution 
suggests a net infall of high-velocity neutral
and weakly ionized gas towards the Milky Way disc.
High ions such as O\,{\sc vi}, in contrast, show
an equal distribution of positive and negative velocities
in the Galactic halo (\cite{2003Sembach}), supporting the idea that
they trace both infalling and outflowing gas that
is highly ionized.

%

\begin{figure}
\centering
\resizebox{1.0\hsize}{!}{\includegraphics[angle=-90]{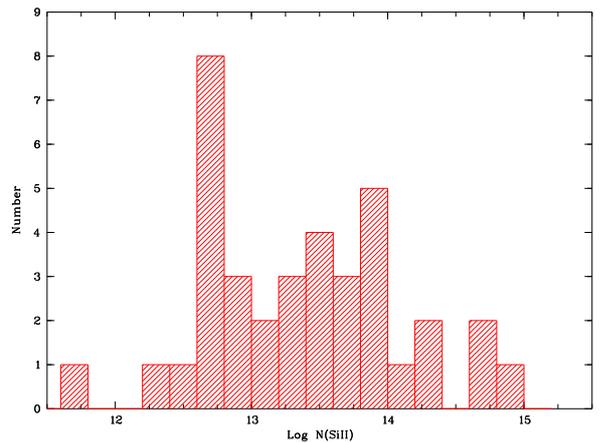}}
\caption[]{Distribution of Si\,{\sc ii} column densities in HVCs, based on
Voigt-profile fitting of 38 HVC absorption components.} 
\label{SiIIhisto}
\end{figure}

%

\subsubsection{Column densities}

From our line-fitting analysis we find that 
the 19 HVCs that are detected and accurately measured 
in Si\,{\sc ii} absorption
are composed of 38 individual absorption components.
For these components we obtain typical Si\,{\sc ii}
column densities in the range log $N$(Si\,{\sc ii}$)=12.5-15.0$ 
(see Tables B.1$-$B.5). A histogram
showing the distribution of the Si\,{\sc ii} column densities
is presented in Fig.\,3. The median logarithmic column density
is log $N$(Si\,{\sc ii}$)=13.48$.

Fig.\,3 indicates a widespread, inhomogeneous
distribution of the Si\,{\sc ii}
column densities in HVC absorption components with 
a prominent peak near log $N$(Si\,{\sc ii}$)\sim 12.8$ and
another maximum near log $N$(Si\,{\sc ii}$)\sim 13.8$.
Most ($21/31$ or $68$ percent) of the HVC absorption components 
have log $N$(Si\,{\sc ii}$)\geq 13.2$.

Tables B.1$-$B.5 show that only some of the Si\,{\sc ii} absorption
components with log $N$(Si\,{\sc ii}$)<13.0$ represent satellite 
components of stronger HVC absorbers.
There exist a distinct population of isolated, weak HVC absorbers 
with relatively low column densities of Si\,{\sc ii} and
other low ions (e.g., towards
PG\,1211$+$143, PKS\,2155$-$304, NGC\,4151). 
These absorbers belong to the class of highly-ionized 
high-velocity clouds (e.g., \cite{1999Sembach}, \cite{2003Sembach}) 
and to the 
low-column-density HVCs (\cite{2009Richter}). This suggests
that the inhomogeneous
distribution in Fig.\,3 reflects the actual
physical properties of neutral and weakly-ionized gas
structures in the Galactic halo, and is not an 
artifact from our analysis.

%

\begin{figure}
\centering
\resizebox{1.0\hsize}{!}{\includegraphics[angle=-90]{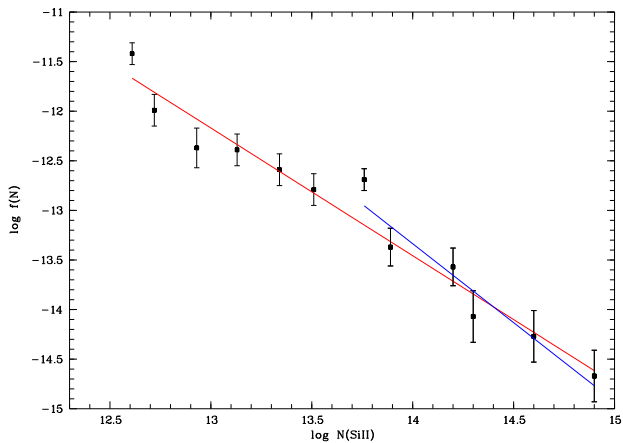}}
\caption[]{Column-density distribution function of Si\,{\sc ii} 
absorption components in HVCs. The data indicate that a 
simple power law (red and blue solid lines) represents a rather poor
approximation to the distribution of Si\,{\sc ii} column
densities.}
\label{CDDF4}
\end{figure}

%

Based on the Si\,{\sc ii} column densities shown
in Fig.\,3 we have constructed a column-density
distribution function (CDDF) of HVC Si\,{\sc ii} absorption
components (Fig.\,4). The CDDF can be defined as
$f(N)=m/\Delta N$, where $m$ is the number of absorbers
in the column-density bin $\Delta N$. The CDDF is usually
approximated by a power law in the form
$f(N)=C\,N^{-\beta}$, where $\beta\approx 1.5$
for H\,{\sc i} in HVCs, as derived from 21cm
observations (e.g., \cite{2002Lockman}).
As can be seen in Fig.\,4, the CDDF of Si\,{\sc ii} HVC
absorption components deviates from a 
simple power law, showing instead a plateau
at log $N$(Si\,{\sc ii}$)\sim 13.8$.
The shape of the CDDF thus reflects the inhomogeneous
distribution of Si\,{\sc ii} column densities in HVC
absorption components shown in Fig.\,3.
If we force a power-law fit with a 
single slope to the Si\,{\sc ii} CDDF for 
column densities log $N$(Si\,{\sc ii}$)\geq 12.5$,
we obtain $\beta=1.29\pm0.09$ and log\,$C=4.57\pm 1.20$
(Fig.\,4, red solid line).
This slope is somewhat shallower than the 
canonical value of $\beta \approx 1.5$ derived from
H\,{\sc i} 21cm observations of Galactic HVCs
(e.g., \cite{2002Lockman}). If we instead
restrict our fit to the range log $N$(Si\,{\sc ii}$)
\geq 13.7$, we obtain much steeper slope
of $\beta=1.59\pm0.23$ and log\,$C=8.94\pm 3.31$
(Fig.\,4, blue solid line).
This slope fits better to the slope derived for
H\,{\sc i}, but is substantially 
smaller than the slope derived
for optical Ca\,{\sc ii} absorption in IVCs and HVCs
($\beta=2.2\pm0.3$; \cite{2008BenBekhti}, \cite{2012BenBekhti}). 
The steeper slope of Ca\,{\sc ii} compared to Si\,{\sc ii}
most likely is a result of the strong depletion of Ca
into dust grains, because high-column density clouds
tend to have higher depletion values than
low-column density clouds (see also 
discussion in \cite{2012BenBekhti}). Note that
if some of the HVC absorption components would be composed 
of several, unresolved velocity components, the
slope of the CDDF would be steeper, too.

\subsubsection{Doppler parameters}

In Fig.\,5 we show the distribution of Si\,{\sc ii} Doppler 
parameters ($b$-values) in HVCs, based on
the Voigt-profile fitting of the 38 Si\,{\sc ii} absorption 
components. The measured $b$-values range from $1$ to $33$
km\,s$^{-1}$ with a median value of $\sim 9.2$ km\,s$^{-1}$.
The distribution can be fitted by a log-normal function
(solid line in Fig.\,5), which peaks at $b=7$ km\,s$^{-1}$.
The median $b$ value is $b=9$ km\,s$^{-1}$.
Note that Si\,{\sc ii} $b$-values that are smaller than the 
instrumental resolution in the E140M grating 
($\sim 7$ km\,s$^{-1}$) can be reliably determined since
we are fitting simultaneously {\it several} Si\,{\sc ii} lines
with different oscillator strengths (i.e., the corresponding
curve-of-growth is well-defined). 

It is commonly assumed that the Doppler parameter
of an absorber is composed of a thermal component
($b_{\rm th}$) and a non-thermal component ($b_{\rm nth}$),
so that $b^2=b_{\rm th}^2+b_{\rm nth}^2$.
The thermal component depends on the 
temperature of the gas, $T$, and the 
atomic weight ($A$) of the absorbing ion:
$b_{\rm th}\approx 0.129\,(T$[K]$/A)^{1/2}$
km\,s$^{-1}$. 
The non-thermal component may include turbulent 
motions in the gas and unresolved velocity structure
in the lines.

Since Si is a relatively heavy element ($A_{\rm Si}=28$),
it is expected that for neutral and partly ionized
HVCs gas with $T\leq 2\times 10^4$ K (e.g., 
\cite{2012BenBekhti}; their Fig.\,15) the thermal contribution
to $b$(Si\,{\sc ii}$)$ is $\leq 3.5$ km\,s$^{-1}$. This
implies that the observed line widths of most of the
Si\,{\sc ii} HVC absorption features are subject to
broadening mechanisms other than thermal broadening,
such as macroscopic turbulence and gas flows.
Moreover, it is very likely that many of the Si\,{\sc ii} 
absorption components seen in the STIS data are composed
of smaller (unresolved) substructures. In fact, optical
absorption-line studies of IVCs and HVCs at very high 
spectral resolution and high S/N clearly indicate that 
there exists substantial velocity-structure in neutral halo clouds
at a level of a few km\,s$^{-1}$ 
(see, e.g., the IVC and HVC in the direction of the Magellanic 
Clouds; \cite{1999Welty}).

%

\begin{figure}
\centering
\resizebox{1.0\hsize}{!}{\includegraphics[angle=-90]{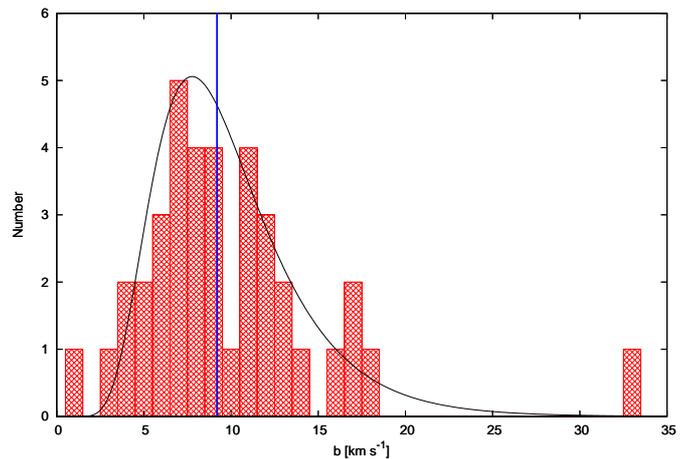}}
\caption[]{Distribution of Si\,{\sc ii} Doppler parameters
in HVCs, based on Voigt-profile fitting of 40 HVC absorption 
components. The black solid line indicates a fit to the distribution
with a log-normal function (see Sect.\,3.2.3). The blue 
solid line marks the median value at 9.2 km\,s$^{-1}$}
\label{b}
\end{figure}

%

\subsubsection{Sub-component structure}

While the smallest substructures in the HVC absorbers 
obviously are not resolved in the STIS data, most of 
the detected Si\,{\sc ii} absorption features do show
several individual velocity subcomponents that are
separated from each other by $>10$ km\,s$^{-1}$,
typically, and that we have fitted as individual
absorption components. Since for intervening
QSO absorbers the velocity spread of the detected
absorption feature often is used as an observational 
parameter to constrain the characteristic environment
of the absorber host (e.g., \cite{1998Charlton}),
it is interesting to study the velocity structure
of Galactic HVCs and compare it to the absorption properties
of intervening systems, in particular weak and strong Mg\,{\sc ii} 
absorbers.

In Fig.\,6 we show the number distribution of Si\,{\sc ii} 
absorption components per HVC for all spectra in which HVC gas 
is detected in Si\,{\sc ii}. About 80 percent of the 
measured HVCs have one or two velocity components that
can be resolved with the STIS data (one component: 32
percent; two components: 47 percent). For comparison,
\cite{2012BenBekhti} find for optical Ca\,{\sc ii} absorption
in IVCs and HVCs that more than 70 percent of the Ca\,{\sc ii} 
absorbers are seen as single-component systems, while the 
fraction of two-component absorbers is less than 20 percent.
This difference is not surprising, however, since Ca\,{\sc ii}
in IVCs and HVCs is expected to trace relatively confined
neutral gas regions in the halo clouds, whereas Si\,{\sc ii}
traces both neutral and weakly ionized gas regions 
(i.e., multi-phase gas regions) that are spatially 
more extended. 

On a first look, the distribution of absorption components in HVCs 
appears to be very similar to the distribution found for 
strong intervening Mg\,{\sc ii} absorbers (\cite{2006Prochter};
one component: $\sim 50$ percent; two components: $\sim 20$
percent).
However, because intervening Mg\,{\sc ii} absorbers trace gas
in discs and halos of galaxies (and thus often are fully
saturated over a large velocity range) and because the spectral
resolution of the Mg\,{\sc ii} SDSS data used 
by \cite{2006Prochter} is very low ($R\approx 2000$), this similarity
does not provide any clues to the connection between strong
Mg\,{\sc ii} systems and HVCs.
Using high-resolution optical spectra, \cite{2003Churchill} indeed 
find a much larger number of $\sim 8$ absorption components
per absorption systems for strong Mg\,{\sc ii} absorbers with
H\,{\sc i} column densities below that expected for neutral gas discs, but
similar to those in Galactic HVCs.

Even if one considers the longer absorption path length through
a galaxy halo from an exterior vantage point (Churchill et al.\,study)
compared to the path length through the Milky Way halo from the 
position of the Sun (our study), the four-times higher number
of absorption components per system clearly indicates that 
the majority of the strong Mg\,{\sc ii} absorbers with
log $N$(H\,{\sc i}$)\leq 20.2$ studied by \cite{2003Churchill} trace
gaseous structures in halos the are kinematically more complex 
than the Galactic HVC population. As we will see later, this scenario is supported
by the very large absorption cross section of strong Mg\,{\sc ii} 
absorbers that are both more common and spatially
more extended (Sect.\,4).

%

\begin{figure}
\centering
\resizebox{1.0\hsize}{!}{\includegraphics[angle=-90]{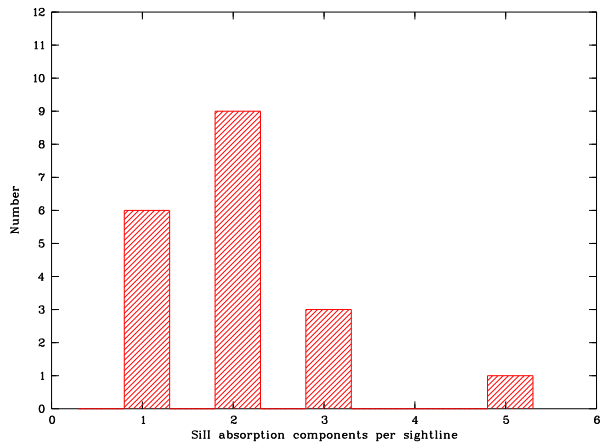}}
\caption[]{Distribution of Si\,{\sc ii} velocity-components in HVCs.}
\label{vLSR1}
\end{figure}

%

\subsection{Remarks on other ions}

\subsubsection{O\,{\sc i}}

O\,{\sc i} is an excellent tracer of H\,{\sc i}, 
because both atoms have the same ionization 
potential and they are coupled by a strong 
charge-exchange reaction. 
There is only one (strong) transition of O\,{\sc i}
available in the STIS E140M wavelength range 
(at $1302.2$ \AA). Thus, $N$(O\,{\sc i}) can
be determined from the STIS data alone, only
under the {\it assumption} that $b$(O\,{\sc i}$)=
b$(Si\,{\sc ii}).

There are several studies on HVCs that have combined
STIS E140M data with FUV data from FUSE (e.g.,
\cite{2001Richter}; \cite{2004Sembach}) to make use of 
several other (weaker) O\,{\sc i} transitions at
$\lambda<1040$ \AA. However, only for a few lines of
sight in our sample there are FUSE data of 
sufficient quality to determine $N$(O\,{\sc i})
at an accuracy similar to that of $N$(Si\,{\sc ii});
in this study, we therefore do not consider any available
FUSE data.

The covering fraction of HVC O\,{\sc i} absorption 
($f_{\rm c}$(O\,{\sc i}$)=0.48$) is smaller
than that of Si\,{\sc ii}, but the column-density limit
above which $f_{\rm c}$(O\,{\sc i}) is considered
is 1.4 dex higher than that of Si\,{\sc ii} (Table 3).
The relative solar abundance of O compared to Si is
log\,(O/Si)$_{\sun}=+1.15$ (\cite{2005Asplund}), 
so that for an HVC with solar relative abundances of
O and Si (and with 100 percent of these elements
in the gas phase) our STIS data are slightly ($0.25$ dex) more
sensitive for Si\,{\sc ii} absorption in HVCs than for
absorption by O\,{\sc i}. On the one hand, the Si\,{\sc ii} 
column density in HVCs may be reduced by the depletion
of Si into dust grains (e.g., \cite{2001Richter};
\cite{2004Richter}); on the other hand, Si\,{\sc ii}
traces neutral {\it and} weakly-ionized gas, so that
Si\,{\sc ii}/O\,{\sc i} may be higher than (O/Si)$_{\sun}$.
Consequently, these effects (dust depletion and ionization) 
are opposite and may even cancel each other out. To estimate whether 
or not this is the case in an HVC one would need to know
the local dust properties and ionization conditions 
in the gas, which are difficult to determine.

The five sightlines, along which high-velocity 
Si\,{\sc ii} is detected without associated O\,{\sc i} absorption
(see Table 2), exhibit relatively weak HVC 
absorption in Si\,{\sc ii}, suggesting that for 
some of these clouds O\,{\sc i} $\lambda 1302.2$ may be 
just below the detection limit (see also \cite{2009Richter}).

\subsubsection{C\,{\sc ii}}

With an ionization potential of $E_{\rm C\,II}=24.4$ eV
singly-ionized carbon traces neutral and mildly ionized
gas in HVCs.
The only available C\,{\sc ii} transition in the 
STIS E140M wavelength range is located 
at $1334.5$ \AA. To derive $N$(C\,{\sc ii})
from the (mostly fully saturated) $\lambda 1334.5$ 
line one needs to assume that $b$(C\,{\sc ii}) is
similar to the Doppler parameter derived for
Si\,{\sc ii} (or other low or intermediate ions). 
However, in view of the higher
ionization potential of C\,{\sc ii} compared
to Si\,{\sc ii}, C\,{\sc ii} absorption may arise
in a somewhat different (possibly more extended)
gas phase, so that this assumption may be invalid
for most of the HVC absorbers. As a consequence, 
the C\,{\sc ii} column densities derived for
our HVC sample are possibly afflicted with large
systematic uncertainties that we cannot account 
for.

From our data we derive a covering fraction of
$f_{\rm c}$(C\,{\sc ii}$)=0.70$; Table 3), which is
insignificantly higher than that of Si\,{\sc ii}.
The column density threshold is log $N_{\rm min}=13.20$,
which is $\sim 1$ dex higher than that of 
Si\,{\sc ii} (see Table 3). This difference 
compensates the expected abundance
difference between these two elements, if
solar relative abundances are assumed
(log (C/Si)$_{\sun}=+0.88$). Therefore,
the C\,{\sc ii} and Si\,{\sc ii} transitions
in the STIS data provide roughly the same 
sensitivity to neutral and weakly
ionized gas in HVCs.

\subsubsection{Mg\,{\sc ii}}

The Mg\,{\sc ii} doublet near 2800 \AA\, is
observed only with the STIS E230M grating, so that
there are only 19 sightlines along which high-velocity
Mg\,{\sc ii} can be studied in our data sample at
intermediate spectral resolution (FWHM$\sim 10$ km\,s$^{-1}$).

As mentioned above, Mg\,{\sc ii} and Si\,{\sc ii} are
expected to trace a similar gas phase in HVCs and both
ions are expected to have very similar column densities.
The Mg\,{\sc ii} filling factor in HVCs
is $f_{\rm c}$(Mg\,{\sc ii}$)=0.53$ (Table 3), which is
somewhat lower than that of Si\,{\sc ii}. This is
not surprising, however, since the Mg\,{\sc ii} column 
density threshold (log $N_{\rm min}=12.70$) 
is $\sim 0.5$ dex higher than that of Si\,{\sc ii}.
In Sect.\,4 we will combine the E230M data for Mg\,{\sc ii} and
the E140M data for Si\,{\sc ii} to compare the absorption
statistics of HVCs with that of intervening
Mg\,{\sc ii} absorbers.

\subsubsection{Fe\,{\sc ii}}

The ionization potential of Fe\,{\sc ii} ($E_{\rm Fe\,II}=16.2$ eV) 
is very similar to that of Si\,{\sc ii} and Mg\,{\sc ii},
so these three ions are expected to arise in the 
same gas phase (neutral and weakly ionized gas).
In the E230M wavelength band there are several 
Fe\,{\sc ii} transitions available including the
two relatively strong transitions at $2382.8$ 
and $2600.2$ \AA. Also the E140M band 
contains a number of (weaker) Fe\,{\sc ii} 
transitions.

The covering fraction of HVC Fe\,{\sc ii} absorption is
$f_{\rm c}$(Fe\,{\sc ii}$)=0.48$ (for log $N_{\rm min}=12.90$), 
thus very similar to that of Mg\,{\sc ii} (see above). 
This is expected, since the threshold column density
is $0.2$ dex higher than for Mg\,{\sc ii}, while
the solar abundance of Fe is $0.08$ dex lower
(log\,(Fe/H$)_{\odot}=-4.55$; \cite{2005Asplund}). 
In summary, both ions 
are equally sensitive to trace predominantly 
neutral and weakly-ionized
gas in HVCs. There are two sightlines that show
Fe\,{\sc ii} absorption at high velocities, while
these two sightlines are not covered in 
Mg\,{\sc ii} (Table 3).

\subsubsection{Si\,{\sc iii}}

Si\,{\sc iii} has one very strong transition at
$1206.500$ \AA; with an ionization potential
of $E_{\rm Si\,III}=33.5$ eV this ion traces
diffuse ionized gas in HVCs and their (more
or less) extended gaseous envelopes. As for 
C\,{\sc ii} and O\,{\sc i} the determination
of a reliable column density for Si\,{\sc iii} 
is basically impossible, as the Si\,{\sc iii} 
$\lambda 1206.500$ line is often heavily 
saturated and the assumption that 
$b$(Si\,{\sc iii}$)=b$(Si\,{\sc ii}) may be 
invalid for most cases.

We find $f_{\rm c}$(Si\,{\sc iii}$)=0.70$, which is
identical to the value derived for C\,{\sc ii}
(i.e., all HVCs that are reliably detected in Si\,{\sc iii}
in our sample are also detected in C\,{\sc ii}).
The column density threshold considered for this
estimate is log $N_{\rm min}$(Si\,{\sc iii}$)=12.15$
(compared to log $N_{\rm min}$(C\,{\sc ii}$)=13.20$;
see above). Since (C/Si)$_{\sun}=+0.88$ (\cite{2005Asplund})
it follows that Si\,{\sc iii} and C\,{\sc ii} trace the
same physical regions in HVCs at roughly the same sensitivity. 
The covering fraction of $f_{\rm c}$(Si\,{\sc iii})$=0.70$
is lower than that derived by Collins et al.\,(2009; 
$f_{\rm c}=0.84$ for log $N_{\rm min}$(Si\,{\sc iii}$)=12.50$)
based on the same STIS data. The reason for this
discrepancy is that these authors also include
very weak (and also spurious) absorption 
features in their statistics that we do not consider
as secure HVC Si\,{\sc iii} detections. In addition,
for our analysis we take into account only those
HVC components that are (in velocity space) 
well separated from lower-velocity material
(i.e., IVCs). Our value of $f_{\rm c}$(Si\,{\sc iii})$=0.70$
is, however, in excellent agreement with the covering
fraction of $f_{\rm c}=0.68\pm 0.04$ of 
UV-selected HVCs based on a much larger STIS/COS data
sample recently presented by \cite{2012Lehner}.

\subsubsection{C\,{\sc iv} and Si\,{\sc iv}}

The high ions C\,{\sc iv} and Si\,{\sc iv} are known to
trace a gas phase in HVCs that is different from that traced
by low ions such as O\,{\sc i}, C\,{\sc ii}, Si\,{\sc ii}, 
Mg\,{\sc ii}, and Fe\,{\sc ii}. High-velocity C\,{\sc iv} and 
Si\,{\sc iv} absorption sometimes is associated with 
common 21cm HVCs (e.g., \cite{2009Fox}), or with low-column
density HVCs (\cite{2009Richter}), where it is thought
to arise in the interface regions between neutral HVC gas 
and the hot coronal gas. In addition, there exists a
population of highly-ionized HVCs (e.g., \cite{1999Sembach};
\cite{2003Sembach}) that probably represent low-density,
gas structures in the halo and that most likely are
photoionized. These structures may arise in diffuse gaseous
material that originates in the IGM and that 
is being accreted by the Milky Way (``warm accretion''),
or that results from the break-up of more massive HVCs 
as they interact with the coronal gas in the halo
(one prominent example is the HVC Complex GCN, which is
detected in C\,{\sc iv} and Si\,{\sc iv} towards
PKS\,2155$-$304 and Mrk\,509 ; see \cite{2011Winkel}).

As covering fractions we derive $f_{\rm c}$(C\,{\sc iv}$)=0.40$
for log $N_{\rm min}$(C\,{\sc iv}$)=13.00$ and 
$f_{\rm c}$(Si\,{\sc iv}$)=0.20$
for log $N_{\rm min}$(Si\,{\sc iv}$)=12.90$. These
covering fractions are smaller than the one derived for 
O\,{\sc vi} in the Milky Way halo ($f_{\rm c}$(O\,{\sc vi})$\geq 0.59$;
\cite{2003Sembach}), implying that O\,{\sc vi} is more
sensitive for detecting highly-ionized halo gas and/or the
O\,{\sc vi} absorbing gas phase is spatially more extended 
than the C\,{\sc iv} and Si\,{\sc iv} absorbing phase.

%

\begin{figure}
\centering
\resizebox{1.0\hsize}{!}{\includegraphics[]{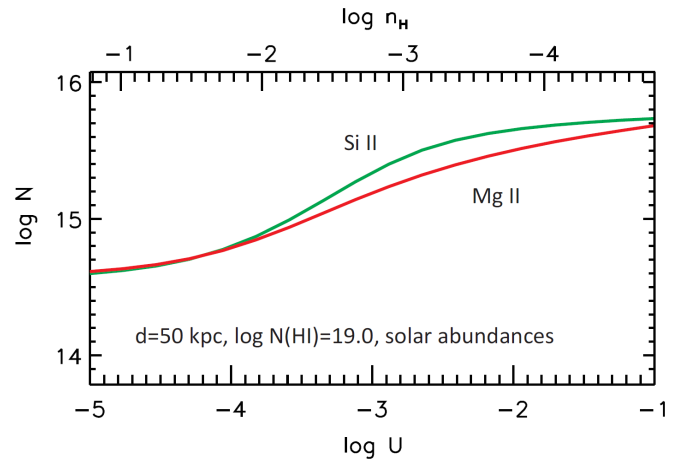}}
\caption[]{Example for a CLOUDY photoionization model of 
Galactic HVC gas at a distance of $D=50$ kpc from the Galactic 
disc, with an H\,{\sc i} column density of log $N$(H\,{\sc i}$)=19$
and a metallicity of $1.0$ solar.
Shown are the expected column densities of Si\,{\sc ii} and Mg\,{\sc ii}
as a function of the gas density, $n_{\rm H}$, and the ionization
parameter, $U$. Because of the similar ionization potentials and
cosmic abundances of both elements/ions the column densities of Si\,{\sc ii} 
and Mg\,{\sc ii} in HVCs are expected to be very similar.}
\label{ionisation}
\end{figure}

%

\section{HVCs as intervening metal-line absorbers}

Deep H\,{\sc i} 21cm observations of M31 and other nearby spiral
galaxies (e.g., NGC\,891) clearly show that the HVC phenomenon
is not restricted to the Milky Way, but represents an ubiquitous
component of spiral galaxies in the local Universe. It indicates
the various gas-circulation processes in the inner and
outer halos of star-forming galaxies (\cite{2004Thilker};
\cite{2007Oosterloo}; \cite{2007Fraternali}; \cite{2012Richter}).
Our results on the covering fraction of the different ions
in the Milky Way HVCs, together with statistics of intervening
metal-line absorbers in QSO spectra,
now can be used to investigate the absorption-cross section
of HVC analogues in the local Universe and to provide an estimate
of the contribution of HVCs to the number density of intervening
metal absorbers at low redshift.

In the following, we will first briefly discuss the general 
relation between the number density of intervening metal absorbers 
and the absorption-cross section of galaxies and their halos. 
As a second step, we will then combine the observed sky covering
fractions of Si\,{\sc ii} and Mg\,{\sc ii} in the Milky Way HVCs
with statistics on intervening Mg\,{\sc ii} absorbers to study the
the spatial distribution of neutral and weakly ionized gas in the halos of 
Milky Way-type galaxies (see also \cite{2011Richter2}; \cite{2012Richter}).

\subsection{Absorption-cross section of galaxies and their halos}

Let us define $d{\cal N}/{dz}\,(X)$ as the absorber 
number density of QSO metal-line aborbers per unit redshift measured
for a given ion $X$, $n_{\rm g}$ as the space density of galaxies at 
$z=0$, $R_{\rm h}(X)$ as the (mean) galaxy halo radius (within 
which metal absorption takes place), and $f_{\rm c}(X)\leq 1$ as the 
mean covering fraction of the ion $X$ in the disc/halo gas 
for radii $r\leq R_{\rm h}(X)$. Number density and geometric
cross section of the gas in and around galaxies are directly
related to each other (e.g., \cite{2008Kacprzak}), so that 
for $z=0$

%

\begin{equation}
\frac{d{\cal N}}{dz}=
\frac{n_{\rm g}\,\langle
f_{\rm c} \rangle \,c\,\pi R_{\rm h}^2}
{H_0}.
\end{equation}

%

The space density of galaxies can also be expressed as
$n_{\rm g}=\Phi^{\star}\,\Gamma(x,y)$, where $\Phi^{\star}$
is the number density of $L^{\star}$ galaxies. $\Gamma(x,y)$
is the incomplete Gamma function in which $x=\alpha +1$, where
$\alpha$ is the slope at the faint end of the Schechter
galaxy luminosity function. The parameter $y$ is defined as
$y=L_{\rm min}/L^{\star}$, where $L_{\rm min}$ is the
minimum luminosity of galaxies contributing to the population
of absorbing gas halos. Therefore, if $n_{\rm g}$ is known 
for a given luminosity range and $d{\cal N}/{dz}\,(X)$ and
$f_{\rm c}(X)\leq 1$ are measured for a given ion, equation
(1) allows us to estimate the characteristic size 
($R_{\rm h}(X)$) of the absorbing region around galaxies.
For a more detailed discussion of these parameters see
\cite{2012Richter}.

\subsection{Covering fraction of strong Mg\,{\sc ii} absorption in the Milky Way halo}

For the study of the physical properties 
and absorption cross section of 
gas in and around galaxies at low and high redshift via
QSO absorption-line spectroscopy the Mg\,{\sc ii} ion
plays a crucial role.
The so-called strong Mg\,{\sc ii} systems 
are intervening metal-absorption systems that have Mg\,{\sc ii}
equivalent widths $W_{2796} > 0.3$ \AA; they usually are associated 
with luminous galaxies ($L>0.05\,L^{\star}$) at impact 
parameters $< 35\,h^{-1}$ kpc (e.g., \cite{1991Bergeron}; 
\cite{2002Steidel}; \cite{2010Kacprzak}).
These absorbers are expected to trace neutral and ionized gas in the
discs of galaxies and their gaeous halos (including HVCs).
The so-called weak Mg\,{\sc ii} systems have $W_{2796} \leq 0.3$ \AA;
they appear to be less tightly associated with galaxies and
are typically found at larger distances from luminous
galaxies, in the range $35-100\,h^{-1}$ kpc 
(\cite{2006Milutinovic}; \cite{2002Rigby}).

From the LCDHVC survey by \cite{2009Richter} and
from this study it is expected that only the most 
massive HVCs (HVCs with neutral column densities
log $N$(H\,{\sc i}$)\geq 17.2$)
display the absorption charcterisitics of strong Mg\,{\sc ii}
absorbers, while there exists a population of HVCs with
H\,{\sc i} column densities log $N$(H\,{\sc i}$)<17.2$
that would appear as weak Mg\,{\sc ii} absorbers if seen
as QSO absorption-line system. The high-velocity
absorbers towards Ton\,S210 and NGC\,4151 represent
examples for this class of low-column density HVCs.
Note that {\it by definition}
all HVCs that have $W_{2796}>0.3$ \AA\, would be classified
as strong Mg\,{\sc ii} systems if seen as QSO absorbers
from far away. However, since all HVCs in
the Milky Way halo have log $N$(H\,{\sc i}$)\leq 20.2$ one would 
expect that HVC analogues in the halos of Milky Way-type galaxies
contribute to the population of strong Mg\,{\sc ii} absorbers
predominantly in the range $0.3$ \AA\,$\leq W_{2796} \leq 1.0$ \AA\,
while the strongest of the intervening Mg\,{\sc ii} absorbers
with $W_{2796}>1.0$ \AA\, are related to discs,
disc-halo interfaces, and galactic winds (e.g., \cite{2012Bouche}).
In the usual QSO absorber classification scheme, HVCs would
appear as sub-damped Lyman $\alpha$ systems (sub-DLAs) and 
Lyman-limit systems (LLS; see \cite{2012Richter}).

In Table 4, sixth row, we show the measured Mg\,{\sc ii} equivalent
widths for the eleven (out of 19 possible) HVCs in which Mg\,{\sc ii} absorption 
has been detected. Strong Mg\,{\sc ii} absorption with $W_{2796}>0.3$ \AA\,
is measured for six HVCs, suggesting that the covering fraction of
HVCs with strong Mg\,{\sc ii} absorption is $f_{\rm c,sMg}=6/19 \approx 0.32$. 
Because of the limited statistical relevance of this result, we also 
consider Si\,{\sc ii} as a proxy for Mg\,{\sc ii}.
As mentioned earlier, Si\,{\sc ii} and Mg\,{\sc ii} have
very similar ionization potentials and solar abundances, so that it is
expected that both ions trace the same gas phase in HVCs and they
have very similar column densites. To demonstrate the expected 
similarity of the Si\,{\sc ii} and Mg\,{\sc ii} column densities 
in HVCs we show in Fig.\,7 an example of a CLOUDY photoionization 
model of a Galactic HVC with an H\,{\sc i} column density of 
$10^{19}$ cm$^{-2}$, a distance to the Galactic disc of 50 kpc,
and with solar abundances of Si and Mg
(model from \cite{2009Richter}). In this figure,
$N$(Si\,{\sc ii}) and $N$(Mg\,{\sc ii}) are plotted against the 
ionization parameter ($U$) and the gas density ($n_{\rm H}$).
Over a large range of densities and ionization parameters
the expected column densities of Si\,{\sc ii} and Mg\,{\sc ii} 
lie within 0.2 dex of each other, demonstrating that these
two ions indeed trace the same gas phase in HVCs with almost identical
column densities.

To transform the measured Si\,{\sc ii} {\it equivalent widths} along the 
19 suited E140M sightlines into Mg\,{\sc ii} equivalent widths, we consider
Si\,{\sc ii} $\lambda 1260$, which is the strongest Si\,{\sc ii} 
transition covered by our STIS data (Table 4, fifth row). 
From atomic data it follows that
($f\lambda)_{\rm Si\,II\,1260}\approx (f\lambda)_{\rm Mg\,II\,2796}$
(\cite{2003Morton}). 

For a typical HVC Doppler parameter range of $b=5-12$ km\,s$^{-1}$
with one or two absorption components
and under the above discussed assumption that $N$(Si\,{\sc ii}$)=N$(Mg\,{\sc ii})
an equivalent width of 300 m\AA\, in the Mg\,{\sc ii} $\lambda 2796$
line corresponds to an equivalent width of $140-200$ m\AA\, in 
the Si\,{\sc ii} $\lambda 1260$ line (see also \cite{2005Narayanan}). 
We here use the lower
threshold of $W_{1260}=140$ m\AA\, to separate strong and weak
Mg\,{\sc ii} absorbers in HVCs in an indirect manner. The observed HVC
Si\,{\sc ii}/Mg\,{\sc ii} absorption strengths towards PG\,1116+215 and 
Ton\,S210 are in excellent agreement with this conversion scheme 
(see Table 4).

Based on this method, we find that 12 of the 20 Si\,{\sc ii} HVC absorbers
listed in Table 4 represent strong Mg\,{\sc ii} systems. 
Combining this result with our direct Mg\,{\sc ii} measurements outlined
above we have 6 relevant Mg\,{\sc ii} plus 8 relevant Si\,{\sc ii} detections 
along 41 independent sightlines, so that 
the total covering fraction of strong 
Mg\,{\sc ii} absorption in Galactic HVCs is estimated as
$f_{\rm c,sMg}=14/41=0.34\pm 0.09$.
This covering fraction for strong Mg\,{\sc ii} absorption 
is identical to the observed covering fraction of H\,{\sc i} 
in HVCs with log $N$(H\,{\sc i}$)\geq 17.8$, based on 21cm HVC surveys
(see \cite{2004Wakker}), and references therein). It is also very similar
to the covering fraction of Ca\,{\sc ii} absorption
in HVCs with log $N$(Ca\,{\sc ii}$)\geq 11.2$, as derived 
from a large sample of optical QSO spectra (\cite{2008BenBekhti}, 
\cite{2012BenBekhti}). Ca\,{\sc ii} traces predominantly 
neutral gas in HVCs with H\,{\sc i} column densities 
log $N$(H\,{\sc i}$)\geq 17.4$ (\cite{2011Richter2};
\cite{2012BenBekhti}). Therefore, the similar absorption
cross sections imply that Ca\,{\sc ii} and strong 
Mg\,{\sc ii} trace the same type of
halo clouds, namely massive HVCs that are optically
thick in H\,{\sc i}; only these clouds would be 
seen as strong Mg\,{\sc ii} absorbers if the Milky Way
halo would be observed as a QSO absorption-line system 
from an exterior vantage point.

%

\begin{table*}[ht]
\caption[]{Equivalent widths of Si\,{\sc ii} and Mg\,{\sc ii} absorption in HVCs.}
\vspace{0.5cm}
\label{EW}
\begin{tabular}{lrrrrr} \hline\hline
Sightline & HVC velocity range &  S/N$_{1260}\,^{\rm a}$   
          & S/N$_{2796}\,^{\rm b}$ 
          & EW$_{\rm SiII,1260}$ & EW$_{\rm MgII,2796}$ \\
          & (km\,s$^{-1}$) &   &  & [m\AA] & [m\AA] \\
\hline
PG\,1211+143    & $+$169...$+$184 & 15   & ...  &  46$\pm$4   & ... \\
PKS\,2155+304   & $-$138...$-$190 & 20   & ...  &  59$\pm$3   & ... \\
Mrk\,279        & $-$159...$-$200 & 23   & ...  &  252$\pm$3  & ... \\
PG\,1116+215    & $+$180...$+$190 & 12   & 6    &  184$\pm$6  & 341$\pm$34 \\
NGC\,3783       & $+$180...$+$250 & 26   & 10   &  356$\pm$4  & 668$\pm$28 \\
NGC\,4151       & $+$120...$+$145 & 18   & 30   &  44$\pm$5   & 43$\pm$3 \\
PKS\,0312$-$770 & $+$160...$+$240 & 7    & 3    &  507$\pm$13 & 1191$\pm$90 \\
HE\,0515$-$4414 & $+$120...$+$230 & ...  & 11   &  ...        & 173$\pm$15 \\
PG\,1634+706    & $-$100...$-$215 & ...  & 25   &  ...        & 1028$\pm$12 \\
NGC\,3516       & $-$160...$-$170 & 3    & 7    & $<200$      & 203$\pm$25 \\
Mrk\,205        & $-$110...$-$230 & 5    & ...  & 253$\pm$18  & ... \\
PHL\,1811       & $-$130...$-$270 & 7    & ...  & 224$\pm$11  & ... \\
PG\,0953+415    & $-$150	  & 8    & ...  & 42$\pm$6    & ... \\
RXJ\,1230.8+0115& $-$216...$-$310 & 6    & ...  & 104$\pm$10  & ... \\
PG\,1259+593    & $-$120...$-$145 & 7    & ...  & 110$\pm$7   & ... \\
Ton\,S210       & $-$150...$-$235 & 6    & 4    & 148$\pm$10  & 305$\pm$6 \\
3c351           & $-$130...$-$230 & 5    & ...  & 389$\pm$17  & ... \\
H\,1821+643     & $-$130...$-$170 & 12   & ...  & 273$\pm$6   & ... \\
NGC\,7469       & $-$190...$-$400 & 10   & ...  & 236$\pm$8   & ... \\
Mrk\,132        & $-$140...$+$80  & ...  & 10   & ...         & 251$\pm$16 \\
PG\,0117+21     & $-$134	  & ...  & 6    & ...         & 83$\pm$21 \\
PG\,1630+377    & $-$160...$-$60  & ...  & 8    & ...         & 653$\pm$30 \\
PG\,1216+069    & $+$210...$+$270 & 8    & ...  & 168$\pm$11  & ... \\
PG\,1444+407    & $-$80...$-$90   & 8    & ...  & 175$\pm$12  & ... \\
3c249           & $-$135	  & 8    & ...  & 27$\pm$11   & ... \\
Mrk 290         & $-$128	  & ...  & ...  & ...         & ... \\
\hline
\end{tabular}
\noindent
\\
$^{\rm a}$ S/N per 3.17 km\,s$^{-1}$ wide pixel element\\
$^{\rm b}$ S/N per 4.83 km\,s$^{-1}$ wide pixel element
\end{table*}

%

\subsection{On the covering fraction of strong Mg\,{\sc ii} 
in the halos of Milky Way-type galaxies}

Being located within the Milky Way disc, we see the distribution
and covering fraction of HVCs in the Galaxy halo 
from the inside-out perspective 
(interior view). For a spherical halo with
radius $R$ the absorption path-length through the halo
is always $\sim R$ and the observed sky-covering fraction,
$f_{\rm c}$, of HVC gas reflects the spatial distribution of gas
integrated from the inner to the outer regions of
the Milky Way halo. If a galaxy and its halo is seen from
an exterior view point, the absorption path length through
the halo (and disc, eventually) depends on the impact
parameter of the sightline, while the observed covering
fraction depends on the path length {\it and} the radial 
gas distribution in the halo.
Therefore, if we want to put into relation the covering fraction 
of Mg\,{\sc ii}/Si\,{\sc ii} in the Milky Way halo with the observed 
number density of intervening Mg\,{\sc ii}/Si\,{\sc ii}
absorbers at low redshift (equation 1), the different 
vantage points need to be considered.
In addition, one needs to consider the absorption cross 
section of gaseous {\it discs}, as intervening absorbers 
passes both disc and halo components of galaxies. This 
important aspect will be discussed in Sect.\,4.4.

To investigate the radial distribution of gas in the halos
of galaxies and the resulting covering fractions, H\,{\sc i} 
21cm studies of nearby galaxies are of crucial importance. 
However, only for a few nearby spiral galaxies (e.g., M31 and NGC\,891; 
\cite{2004Thilker}; \cite{2007Oosterloo}) are the 21cm observations 
deep enough to provide meaningful constraints on the 
distribution of neutral gas in their halos.
In a recent study, \cite{2012Richter} has demonstrated that the projected 
covering fraction of 21cm HVCs around M31 strongly decreases
with radius and can be fitted by an exponential in the form

%

\begin{equation}
f_{\rm HVC}=2.1\,{\rm exp}\,(-r/h), 
\end{equation}

%

where $r$ is the projected radius in [kpc] and $h=12$ kpc is the
scale height for H\,{\sc i} in HVCs. 
An exponential decline of the covering fraction of
neutral/ionized gas in the halos of Milky Way-type galaxies is
further supported by high-resolution hydrodynamical simulations of
galaxies (\cite{2012Fernandez}).

\cite{2012Richter} developed the numerical code {\tt halopath} that can 
be used to calculate the covering fraction of neutral and ionized halo 
gas from any given vantage point inside and outside the halo sphere.
The {\tt halopath} code assumes that the neutral and ionized 
gas in galaxy halos is distributed spherically around
the neutral gas discs of these galaxies. Instead of modeling 
individual halo clouds or halo-gas structures (which would require 
knowledge about the size distribution of such structures), the code
uses the {\it volume} filling factor of a given gas phase (e.g., 
neutral or ionized gas) as a function of galactocentric
distance, $f_{\rm v}(R)$, as main input parameter. The function
$f_{\rm v}(R)$ can be parametrized for an individual galaxy or for
a population of galaxies. A corresponding model for the gas discs
can also be included. The code then delivers the 
absorption cross section for each gas phase as a function
of galaxy impact parameter and the total area covered by gas in the discs
and halos of the modeled galaxies from an exterior vantage point.
The code also calculates the sky covering fraction of each gas phase
if a vantage point {\it inside} the sphere is chosen; it thus allows
us to link the distribution of Milky Way HVCs with the frequency
of intervening QSO absorbers. For more details on the code and 
its application to gaseous galaxy halos see \cite{2012Richter}.
One important result from the study by \cite{2012Richter} is that the observed
sky covering fraction of HVCs in the Milky Way halo (interior view)
is fully consistent with the projected covering fraction 
of H\,{\sc i} clouds in the halo of M31 (exterior view). This
suggests that the distribution of neutral gas in the halos
of both galaxies is similar (in a statistical sense) and that
basically all HVCs lie within $50$ kpc from the discs.

We now use the code {\tt halopath} to estimate the total
absorption cross section and mean covering fraction of strong 
Mg\,{\sc ii} in HVC analogues around Milky Way/M31 type galaxies 
from an external vantage point.
As model input we adopt the observed sky covering fraction of
$f_{\rm c,sMgII,halo,i}=0.34$ of strong Mg\,{\sc ii} in
the Milky Way from the {\it interior} vantage point.
We further assume that the observed 
sky covering fraction of strong Mg\,{\sc ii} absorption in the 
Milky Way HVCs (interior view) is representative for non-star forming
disc galaxies of similar mass and further assume that the 
projected covering fraction of halo Mg\,{\sc ii} 
(exterior view) declines expontially (i.e., $f_{\rm MgII}(r)$ 
declines in same way as $f_{\rm HI}(r)$; see equation 2).

Under these assumptions, we find (using the {\tt halopath} code)
that the projected covering fraction of strong Mg\,{\sc ii}
from HVCs in the halos of Milky Way-type galaxies 
as seen from an external vantage point is 
$\langle f_{\rm c,sMgII}\rangle \approx0.2$ for a halo radius of
$r_3=61$ kpc 
\footnote{We define $r_3$ as the halo radius beyond which the projected
covering fraction of strong Mg\,{\sc ii} falls
below the 3 percent level.}. The projected halo area covered
by strong Mg\,{\sc ii} in HVCs around Milky Way-type 
galaxies then turns out to be $A_{\rm sMgII,halo}=2340$ kpc$^2$.

\subsection{On the contribution of HVCs to the absorber density of 
strong Mg\,{\sc ii} absorbers}

Our STIS measurements imply that HVCs and their distant analogues
have a non-negligable absorption cross section in low- and 
intermediate ions such as Si\,{\sc ii} and Mg\,{\sc ii}, and thus
these objects are expected to contribute to the observed number density 
of strong Mg\,{\sc ii} absorbers. However, for a 
quantitative estimate of the contribution of HVCs to 
$d{\cal N}/dz$(Mg\,{\sc ii}) one also needs to consider the 
the absorption cross section of gaseous {\it discs},
as intervening absorbers passes both disc and halo components
of galaxies. 
While a detailed discussion of the absorption cross section of
gaseous galaxy discs is beyond the scope of this paper, we
provide some simple estimates that help to evaluate the
relevance of HVCs for the absorption cross section of 
strong Mg\,{\sc ii} absorbers.

For the Mg\,{\sc ii}-absorbing disc component in our Milky
Way/M31 model galaxy we assume a radius of $r_{\rm disc}=30$ kpc
(for log $N$(H\,{\sc i}$)>17.5$), based on the M31 21cm data
of \cite{2009Braun}, and a covering fraction
of $f_{\rm c,sMgII,disc}=1$. The mean absorption cross section for
strong Mg\,{\sc ii} of a sample of randomly inclined gas 
discs with these properties 
then is $A_{\rm disc}\approx 1810$ kpc$^2$, which is $\sim 77$ percent
of the cross section of the surrounding HVC population (see previous
section). Since the areas covered by discs and halo clouds are 
overlapping from an exterior vantage point, projection effects 
need to be taken into account. Using the {\tt halopath} code
we calculate that the absorption cross section of strong Mg\,{\sc ii}
of gas discs plus HVCs is $A_{\rm disc+HVC}\approx 3790$ kpc$^2$,
thus a factor of $\sim 2.1$ higher than the cross section of gas discs 
{\it without} a population of surrounding Milky-Way type HVCs.
Our calculations therefore imply that, from an exterior 
vantage point, $\sim 52$ percent 
($f=1980/3790=0.522$) of the cross section of strong Mg\,{\sc ii}
absorption in the Milky Way and M31 comes from the HVC
population in their halos. Finally, the mean covering fraction of strong 
Mg\,{\sc ii} from the disc plus HVCs in our model halo with radius 
$r_3=61$ kpc, as seen from an exterior vantage point, is 
$\langle f_{\rm c,sMgII}\rangle =0.31$.

While the Milky Way and M31 show very similar properties in
their HVC populations, it is unclear whether these two galaxies
are representative for the local galaxy population
with respect to their halo-gas distribution. 
Nearby spiral galaxies do show extended H\,{\sc i} halos
and extraplanar gas features that may be regarded as distant
HVC analogues (\cite{2007Oosterloo}; 
\cite{2008Sancisi}), but with only few galaxies observed at sufficient 
21cm sensitivity no meaningful conclusions can be drawn. 
Yet, coherent H\,{\sc i} structures at relatively large galactocentric 
distances $d>30$ kpc, such as tidal features like the Magellanic Stream (MS),
appear to be rare. Note that the MS accounts for a 
substantial fraction of the neutral gas mass and H\,{\sc i} covering fraction 
in the Milky Way halo (see \cite{2012Richter}; their Table 1). 
Many of the high-velocity absorption features in the southern sky (Fig.\,1)
may be associated with gas originating in the Magellanic Stream. Therefore,
the absorption cross section of gaseous halos of galaxies, that are not
surrounded by tidal gas streams, could be substantially smaller. 
Our STIS sample is not large enough to investigate the absorption cross
section of the Milky Way HVCs in different quadrants of the Galaxy 
on a statistically secure basis to further study this effect in detail. 
However, using the quickly growing COS data archive, we will address this 
interesting aspect in a subsequent paper.

Even if one assumes that Milky Way-type HVCs (including
the MS) are common in low-redshift galaxies, they cannot
dominate the number density of strong Mg\,{\sc ii} absorbers
at $z=0$: \cite{2012Richter} estimate that the contribution of
Milky-Way type HVCs to the number density of intervening absorbers is
$d{\cal N}/dz<0.167$, assuming that the HVC distribution 
around the Milky Way and M31 is typical for the local galaxy
population. This number density is $<34$ percent of the 
expected value for intervening
strong Mg\,{\sc ii} absorbers at $z=0$ 
($d{\cal N}/dz$(Mg\,{\sc ii}$)\approx0.5$; \cite{2005Nestor}).

The contribution of gas discs to the Mg\,{\sc ii} 
cross section is expected to be even smaller.
From 21cm studies of the H\,{\sc i} mass function of
nearby galaxies (e.g., \cite{2005Zwaan}) follows
that neutral gas discs with log $N$(H\,{\sc i}$)\geq 20.3$ contribute
with only $d{\cal N}/dz\approx0.045$ to the local population of
strong Mg\,{\sc ii} absorbers, a number density that is one order of magnitude
smaller than that of strong Mg\,{\sc ii} systems. At this
H\,{\sc i} column density limit, intervening absorbers would
be identified as Damped Ly\,$\alpha$ systems (DLAs) in the
common classification scheme for QSO absorption-line systems.
One may argue that the radial extent and cross section 
of strong Mg\,{\sc ii} absorbing gas discs extends is much 
larger for column densities {\it below} the DLA column density
limit. There is, however, no compelling observational evidence 
for the presence of extended H\,{\sc i} discs beyond the DLA
column density limit. On the contrary,
recent 21cm studies of the radial extent of the H\,{\sc i} discs
of low-redshift galaxies indicate that gas discs typically
are radially truncated with a sharp drop in H\,{\sc i}
column density at the disc edge from $10^{20}$ cm$^{-2}$ down to $10^{17}$
cm$^{-2}$ (\cite{2009Braun}; \cite{2007Oosterloo}; \cite{2009Portas}).
From the composite radial H\,{\sc i} disc profile of the
galaxies in the THINGS survey (\cite{2009Portas}; their Fig.\,5)
it follows that the disc radius (disc area) for log $N$(H\,{\sc i}$)>17$ is
only $\sim 10$ ($\sim 21$) percent larger than for 
log $N$(H\,{\sc i}$)>20$, so that the contribution of extended 
gas discs to the observend number density of intervening absorbers is 
is limited to $d{\cal N}/dz\approx0.054$, which is $<11$ percent
of $d{\cal N}/dz$(Mg\,{\sc ii}) at $z=0$.

In conclusion, neither the discs of Milky Way-type galaxies nor the 
HVCs in their halos provide enough cross section in Mg\,{\sc ii} 
to explain the large number density of intervening strong Mg\,{\sc ii} 
absorbers. Other phenomena that transport neutral and ionized gas 
outside of galaxies appear to dominate the absorption cross
section of Mg\,{\sc ii} in the local Universe.
This conclusion is supported by {\it direct} estimates of 
the absorption cross section of Mg\,{\sc ii} absorbing
galaxies. For intermediate redshifts, \cite{2008Kacprzak} derived 
a larger mean covering fraction of 
$\langle f_{\rm c,sMgII} \rangle \approx 0.5$ and 
larger halo radii for Milky-Way size galaxies in 
their galaxy sample. Based on observations
at low $z$, \cite{2012Bouche}
argued that galaxy outflows and winds must dominate the 
total cross section of strong Mg\,{\sc ii} absorption
in the local Universe. This, in turn, is in line with
our scenario that 
``quiescent'' galaxies (i.e., galaxies that do not
drive extended winds and outflows), such as the Milky Way and M31,
contribute relatively little to the observed number density
of strong Mg\,{\sc ii} absorption at low redshift.
Moreover, Mg\,{\sc ii} absorption in
the halos of such galaxies predominantly indicates material that
is {\it infalling} rather than outflowing. 

\subsection{On the contribution of HVCs to the absorber density of
weak Mg\,{\sc ii} absorbers}

Intervening Mg\,{\sc ii} absorbers with equivalent widths $<300$ m\AA\, 
in the Mg\,{\sc ii} $\lambda 2796$ line are commonly referred to as
weak Mg\,{\sc ii} absorbers. With a number density of $d{\cal N}/dz=1.00$
at $z<0.3$ for rest-frame equivalent widths $W_{2796}=20-300$ m\AA\, 
these systems have a similarly high absorption cross section
as the strong Mg\,{\sc ii} absorbers (\cite{2005Narayanan}).
However, compared to strong Mg\,{\sc ii} systems, weak Mg\,{\sc ii} 
absorbers probably are located at larger distances from the 
galaxies, i.e., in their circumgalactic environment
(\cite{2006Milutinovic}). 

\cite{2009Richter} have speculated that some of the weak high-velocity
O\,{\sc i}/Si\,{\sc ii}/Mg\,{\sc ii} absorbers in the STIS
spectra, that have no associated H\,{\sc i} 21cm emission,
may be located at larger distances ($d>60$ kpc) than the more
massive 21cm HVCs and thus possibly 
represent the local analogues of the intervening weak Mg\,{\sc ii}
absorbers found at low and high redshift. From our study we
infer a covering fraction of $f_{\rm c,wMg}=10/41\approx 0.24$
for weak Mg\,{\sc ii} absorption at high-velocities for 
equivalent widths $W_{2796}=40-300$ m\AA; this covering fraction 
is slightly lower than that of strong Mg\,{\sc ii} absorption.

For the same line of arguments outlined in the previous section,
the contribution of the 21cm HVCs at $d<60$ kpc to the number
density of weak Mg\,{\sc ii} absorbers must be small. If, however,
low-column density Mg\,{\sc ii}/Si\,{\sc ii} absorbers in
the Milky Way halo were typically located within $R_{\rm h}$ 
at distances $d>60$ kpc (as proposed by \cite{2009Richter}), 
then the absorption cross section of such objects from an exterior 
vantage point and their contribution to $d{\cal N}/dz$
would be increased substantially 
(since $d{\cal N}/dz \propto A \propto R_{\rm h}^2$; equation 1).
The ongoing COS observations of weak Mg\,{\sc ii}/Si\,{\sc ii} 
absorbers along sightlines that pass the inner and outer halos of 
nearby galaxies (e.g., \cite{2011Ribaudo2}) 
will help to further explore this scenario.
 
%

\section{Summary}

In this paper, we have used archival UV absorption-line data
from \emph{HST}/STIS to statistically analyse the absorption 
properties of metal ions such as 
O\,{\sc i}, C\,{\sc ii}, Si\,{\sc ii}, Mg\,{\sc ii}, 
Fe\,{\sc ii}, Si\,{\sc iii}, C\,{\sc iv}, and Si\,{\sc iv} 
in high-velocity clouds (HVCs) in the 
Galactic halo towards more than 40 extragalactic background
sources. Our main results are the following:\\
\\
1) UV absorption features related to HVC gas is detected along 28
sightlines out of 47. The absorption covering fraction, $f_{\rm c}$, 
for the different ions vary between 0.20 (Si\,{\sc iv}) and 0.70
(C\,{\sc ii}, Si\,{\sc iii}) for the column density thresholds
chosen for our data (Table 2).
\\
\\
2) We identify Si\,{\sc ii} with its five detectable transitions 
in the STIS wavelength band as the best-suited ion to statistically 
study the UV absorption characterisitics of neutral and weakly ionized 
gas in HVCs.
We find that the absorption covering fraction of Si\,{\sc ii} in HVC
gas is $f_{\rm c}$(Si\,{\sc ii}$)=0.67$ for log $N$(Si\,{\sc ii}$)
\geq 12.25$. This is a factor of two higher than the filling factor
of H\,{\sc i} in HVCs for log $N$(H\,{\sc i}$) \geq 17.8$. Therefore,
Si\,{\sc ii} is a sensitive tracer for neutral and partly ionized halo 
gas with H\,{\sc i} column densities below the detection limit of 
H\,{\sc i} 21cm HVC  surveys.
\\
\\
3) About 70 percent of the high-velocity Si\,{\sc ii} absorption is detected
at negative radial velocities (albeit at relatively low significance), 
pointing towards a net infall of neutral
and weakly ionized gas traced by Si\,{\sc ii}. Most of these
HVC absorbers are composed of $1-2$ velocity subcomponents that
have Doppler-parameters in the range $b=5-15$ km\,s$^{-1}$. 
Non-thermal line broadening mechanisms dominate the observed
Si\,{\sc ii} $b$-values in HVCs.
\\
\\
4) For the 38 individual Si\,{\sc ii} absorption components in HVCs
we obtain typical column densities in the range log $N$(Si\,{\sc ii})
$=12.5-15.0$. The Si\,{\sc ii} column densities show an irregular
distribution with an apparent deficiency of absorbers with
log $N$(Si\,{\sc ii}$)\approx 13$. As a result, the column-density
distribution function of HVC Si\,{\sc ii} absorption
components is not well described by a single-slope 
power law in the form $f(N)=C\,N^{-\beta}$. If we,
however, force a fit to such a power law we obtain 
$\beta=1.34\pm0.12$ and log\,$C=5.21\pm 1.72$. This slope
is mildly shallower than the slope of $\beta \approx 1.5$ 
for H\,{\sc i} in HVCs obtained from 21cm observtions.
\\
\\
5) Because Si\,{\sc ii} and Mg\,{\sc ii} have similar ionization
potentials and both elements have similar cosmic abundances, we 
combine the information from Si\,{\sc ii} and Mg\,{\sc ii} absorption
in the Galactic HVCs to investigate the covering fraction of 
strong Mg\,{\sc ii} absorption (with equivalent widths $W_{2796} > 0.3$ \AA) 
in the Milky Way halo. We find that the covering fraction of 
strong Mg\,{\sc ii} absorption in the Galactic halo is
$f_{\rm c,sMgII,halo,i}=0.34\pm 0.09$ from our position within the 
Milky Way disc. This value is similar to the covering fraction
of H\,{\sc i} in HVCs at column densities log $N$(H\,{\sc i}$)\geq 17.8$.
Our analysis implies that only the most massive Galactic HVCs would represent 
strong Mg\,{\sc ii} absorbers if seen as intervening absorbers 
from an external vantage point.
\\
\\
6) We combine our results with the geometrical
Milky Way/M31 halo model 
by \cite{2012Richter} to estimate the cross section of 
strong Mg\,{\sc ii} absorption in the Milky Way HVCs if
they would be seen as a QSO absorber from an exterior vantage point.
For disc and halo components together we obtain a 
mean covering fraction of strong Mg\,{\sc ii} absorption
of $\langle f_{\rm c,sMgII}\rangle =0.31$ for a halo radius 
of $R=61$ kpc. From this follows that $\sim 52$ percent
of the cross section of strong Mg\,{\sc ii}
absorption in the Milky Way comes from the HVC
population in the Galactic halo.
\\
\\
7) Our results, together with the \cite{2012Richter} HVC model,
indicate that the contribution of HVCs to the number density
($d{\cal N}/dz$) of strong Mg\,{\sc ii} absorbers at $z=0$
is small, but not negligable ($< 34$ percent). 
These findings are in line with the idea that most of the
intervening strong Mg\,{\sc ii} absorbers are related
to gaseous outflows and galactic winds arising in the extended 
halos of more actively star-forming galaxies.
\\
\\
For the future, we are planning to continue our studies on the
UV absorption chararacteristics of Galactic HVCs and their relation
to intervening QSO absorbers using the \emph{HST}/COS data archive
that is quickly filling with fresh UV absorption-line spectra of
low-redshift QSOs.

%

\begin{acknowledgements}

P.H. and P.R. acknowledge financial support by the German
\emph{Deut\-sche For\-schungs\-ge\-mein\-schaft}, DFG,
through grant Ri\,1124/$8-1$. 

\end{acknowledgements}

%

\bibliographystyle{aa}
\bibliography{bibfile}

%

\appendix

\section{Discussion on individual HVC sightlines}

In the following we shortly discuss the HVC absorption properties
individually for each line of sight and summarize the results from
previous studies. The sightlines are sorted by Galactic
longitude.

{\it\bf PKS\,2155$-$304.}
The PKS\,2155$-$304 sightline is located at $l=17.7$, $b=-52.3$;
it thus passes the outer edge of HVC Complex GCN. HVC absorption
is seen at high negative velocities. Only E140M
data are available for this sightline. The data have good
quality with a S/N of $\sim 18$ per $3.2$ km\,s$^{-1}$-wide
pixel element at 1300 \AA. HVC absorption is split into
two groups (see Fig.\,B.4; Table B.4).
The first group near $v_{\rm LSR}=-150$ km\,s$^{-1}$
is detected in C\,{\sc ii}, Si\,{\sc ii}, Si\,{\sc iii},
Si\,{\sc iv} and C\,{\sc iv} and has three subcomponents
at $-111, -135$ and $-157$ km\,s$^{-1}$. The second group
is centred near $-260$ km\,s$^{-1}$ and is detected
only in Si\,{\sc iii}, Si\,{\sc iv} and C\,{\sc iv}.
It has three subcomponents at $-232, -254$ and $-280$ km\,s$^{-1}$.
Detailed analyses of the PKS\,2155$-$304 sightline are
presented by \cite{1999Sembach} and \cite{2004Collins}. 
This sightline is also discussed in \cite{2003Sembach},
\cite{2006Fox}, \cite{2009Collins}, and
\cite{2011Winkel}.

{\it\bf Mrk\,509.}
The Mrk\,509 sightline is located at $l=36.0$, $b=-29.9$ and passes
through highly-ionized HVC gas at very high negative velocities.
This HVC is associated with Complex GCN (see \cite{2011Winkel}).
Both E140M and E230M data are available for this sightline,
but the S/N in the data is relatively low ($\sim 7 $ per
$3.2$ km\,s$^{-1}$-wide pixel element at 1300 \AA). Absorption
is detected in four components at $v_{\rm LSR}=-263,
-287, -273$, and $-311$ km\,s$^{-1}$ in the lines of
C\,{\sc ii}, Si\,{\sc iii}, Si\,{\sc iv} and C\,{\sc iv}
(see Fig.\,B.2; Table B.5). Detailed studies of this
HVC are presented in \cite{2004Sembach} and
\cite{2004Collins}. Other studies that
discuss this sightline are \cite{2003Sembach}
and \cite{2011Winkel}.

{\it\bf PHL\,1811.}
The line of sight towards the Seyfert 1 galaxy PHL\,1811
($l=47.5$, $b=-44.8$) passes the outskirts of HVC Complex
GCN and shows a complex HVC absorption pattern in the
LSR veocity range between $-100$ and $-300$ km\,s$^{-1}$
(see Fig.\,B.3; Table B.2). Only E140M data are available
for this sightline; the S/N is $\sim 7$
per $3.2$ km\,s$^{-1}$-wide pixel element at 1300 \AA.
Five absorption components are identified at
velocities of $v_{\rm LSR}=-163,-206,-240,-263$, and
$-351$ km\,s$^{-1}$ in the lines of C\,{\sc ii},
O\,{\sc i}, Si\,{\sc ii}, Al\,{\sc ii}, Si\,{\sc iii},
Si\,{\sc iv}, and C\,{\sc iv}. The large number
of components suggests a complex spatial distribution
of HVC gas in Complex GCN in this direction (see also
\cite{2011Winkel}). A detailed study of the Complex GCN
absorption towards PHL\,1811 is presented in 
\cite{2009Richter}. Other studies of relevance in this context
are the ones by \cite{2006Fox}, \cite{2008BenBekhti}, \cite{2009Collins}, and
\cite{2011Winkel}.

{\it\bf PG\,1630$+$377.}
For the line of sight towards the quasar PG\,1630$+$377 ($l=60.3$, $b=+42.9$) only 
STIS data from the E230M grating are available (see Fig.\,B.3; Table\,B.3). 
HVC absorption is detected in Mg\,{\sc ii} and Fe\,{\sc ii} in 
three individual components 
at negative velocities at $v_{\rm LSR}=-64, -99$, and $-155$ km\,s$^{-1}$. 
With a S/N of $\sim$8 per $4.8$ km\,s$^{-1}$ wide pixel element at 2796 \AA\,the 
data quality is rather low. For the two absorption components 
near $-100$ km\,s$^{-1}$ the Mg\,{\sc ii} absorption is saturated. 
We did not find any previous studies that discuss HVC absorption along this
line of sight.

{\it\bf PG\,1444$+$407.}
Along the line of sight towards the Seyfert\,1 galaxy PG\,1444$+$407 at 
$l=69.9$ and $b=+62.7$ we identify one double-component high-velocity absorber
at $v_{\rm LSR}\approx -85$ km\,s$^{-1}$ in the lines of Si\,{\sc ii}, 
O\,{\sc i}, C\,{\sc ii}, and C\,{\sc iv} (see Fig.\,B.3; Table\,B.3). 
The C\,{\sc iv} component seems to be shifted to more negative velocities 
and the C\,{\sc ii} absorption is strongly saturated. This indicates 
a two-phase HVC structure with an inner core that is surrounded by an ionized
envelope. The data have a S/N$\sim 8$ per $3.2$ km\,s$^{-1}$-wide pixel 
element at 1260 \AA. This HVC absorber is located at the outer edge of 
Complex C. For further information see the sample of \cite{2011Wakker}
and \cite{2009Shull}.

{\it\bf PG\,1718$+$481.}
The line of sight towards the QSO PG\,1718$+$481 is located
at $l=74.4$, $b=+34.8$; it therefore pierces the outer environments
of the HVC Complexes C and K. Unfortunately, only Fe\,{\sc ii}
$\lambda 2600$ is covered by the E230M data, while E140 data
are not available. The E230M data have a good S/N of $\sim 14$
per $4.8$ km\,s$^{-1}$-wide pixel element at 2600 \AA. The
Fe\,{\sc ii} line shows three well-defined HVC components
at radial velocities of $v_{\rm LSR}=-81, -128$, and $-197$
km\,s$^{-1}$ (see Fig.\,B.3; Table B.2). We did not find any
previous paper that discusses the HVC absorption towards
PG\,1718$+$481.

{\it\bf NGC\,7469.}
The line of sight towards NGC\,7469 passes directly through
the Magellanic Stream at $l=83.1$, $b=-45.5$. Consequently, the
STIS data show a complex absorption pattern at high negative
velocities in the range $v_{\rm LSR}=-180$ to $-400$ km\,s$^{-1}$.
For this sightline only E140M data are available. The
S/N in the data is $\sim 10$ per $3.2$ km\,s$^{-1}$-wide pixel element
at 1300 \AA. We fit five individual absorption components to
the data at $v_{\rm LSR}=-185,-251,-293,-335$, and $-366$ km\,s$^{-1}$
(Fig.\,B.2; Table B.4). A detailed analysis of this sightline
is presented in \cite{2010Fox}.

{\it\bf 3C\,351.}
The 3C\,351 sightline passes the HVC Complexes C
and K at $l=90.1$, $b=+36.4$. Only E140 M data
are available for this sightline. The S/N
in the data is relatively low ($\sim 6$ per
$3.2$ km\,s$^{-1}$-wide pixel element at 1500 \AA).
A complex absorption pattern at high negative velocities
is visible in the absorption lines of
C\,{\sc ii}, Al\,{\sc ii}, Si\,{\sc ii}, O\,{\sc i},
Si\,{\sc iii}, Si\,{\sc iv}, and C\,{\sc iv} (see Fig.\,B.1;
Table B.1). We identify five absorption components
at LSR velocities of $-76$, $-89$, $-131$, $-166$, and
$-192$ km\,s$^{-1}$. HVC absorption in Complex C and Complex K
has been studied in detail by \cite{2003Tripp} and
\cite{2007Collins}.

{\it\bf Mrk\,290.}
The line of sight towards Mrk 290 ($l=91.5$, $b=+48.0$) is known 
to pass through Complex C. Unfortunately there is only E230M data 
available to us, and only the two Fe\,{\sc ii} lines are covered. 
We find HVC absorption near $v_{\rm LSR}=-130$km\,s$^{-1}$ in three
absorption components, but because of the low S/N 
($\sim 8$ per $4.8$ km\,s$^{-1}$-wide
pixel element at 2600 \AA) and the saturation of the Fe\,{\sc ii} lines 
a reliable measurement is not possible (see Fig.\,B.2;
Table B.5). HVC absorption along this sightline is discussed 
also in \cite{2011Shull}.

{\it\bf H1821$+$643.}
The H1821$+$643 sightline at $l=94.0$, $b=+27.4$ is known to
pass through the Outer Arm and the HVC Complexes C and K.
Thus, the available E140M STIS spectrum shows a complex
absorption pattern of high-velocity gas components in
the velocity range $-80$ to $-200$ km\,s$^{-1}$. The E140M data
are of good quality with a S/N of $\sim 11$
per $3.2$ km\,s$^{-1}$-wide pixel element at 1300 \AA.
We identify three distinct absorption components
at $v_{\rm LSR}=-84,-126$, and
$-146$ km\,s$^{-1}$ in the lines of C\,{\sc ii},
O\,{\sc i}, Si\,{\sc ii}, Al\,{\sc ii}, Si\,{\sc iii},
Si\,{\sc iv}, C\,{\sc iv}, and others (see Fig.\,B.1;
Table B.2). A detailed analysis
of the HVCs towards H1821$+$643 is presented in
\cite{2003Tripp}. Other studies that discuss HVC
absorption along this sightline are the ones by
\cite{2003Sembach} and \cite{2009Collins}.

{\it\bf PG\,1634$+$706.}
The PG\,1634$+$706 sightline passes HVC Complex C at
$l=102.8$, $b=+36.6$. Only E230M data are available for
this sightline, so that
Complex C absorption is detected only in the lines of
Fe\,{\sc ii} and Mg\,{\sc ii} at LSR velocities in the range
$-100$ to $-230$ km\,s$^{-1}$ (see Fig.\,B.3; Table B.2).
The E230M data have a high S/N of $\sim 22$ per $4.8$ km\,s$^{-1}$-wide
pixel element at 2800 \AA.
Three major absorption components are identified at
$v_{\rm LSR}=-125, -164$, and $-195$ km\,s$^{-1}$. The
latter component appears to be very broad in Mg\,{\sc ii}
and possibly is composed of several (unresolved)
sub-components. We did not find any previous study in the
literature that analyses the HVC absorption in the STIS E230M data
towards PG\,1634$+$706.

{\it\bf Mrk\,279.}
The line of sight towards Mrk\,279 pierces HVC Complex C
at $l=115.0$, $b=+46.9$. The available E140M data show
strong HVC absorption at negative velocities in the range
$v_{\rm LSR}=-100$ to $-200$ km\,s$^{-1}$ in the lines
of C\,{\sc ii}, Si\,{\sc ii}, O\,{\sc i}, Al\,{\sc ii},
Si\,{\sc iii} and Si\,{\sc iv}. The data have good quality
with a S/N of $\sim 14 $ per $3.2$ km\,s$^{-1}$-wide pixel element
at 1500 \AA. Three individual components at $v_{\rm LSR}=-145,
-161$, and $-179$ km\,s$^{-1}$ are fitted to the data
(see Fig.\,B.1; Table B.5). There exist a number
of detailed studies of the Complex C absorption
towards Mrk\,279 based on different spectral data
(e.g., \cite{2001Gibson}; \cite{2003Tripp}; 
\cite{2003Collins}, \cite{2007Collins}).

{\it\bf PG\,1259$+$593.}
The line of sight towards the QSO PG\,1259$+$593 ($l=120.6$, $b=+58.0$)
passes HVC Complex C at negative LSR velocities; it
represents one of the best-studied HVC
sightlines in the literature. Because PG\,1259$+$593 is relatively faint
($V=15.84$ mag) the existing E140M data (obtained with
more than $350$ ksec integration time; see \cite{2003Sembach})
has only an intermediate S/N of $\sim 7$ per $3.2$ km\,s$^{-1}$-wide
pixel element at 1500 \AA. Strong HVC absorption associated with
Complex C is seen in two distinct absorption components at
$v_{\rm LSR}=-134$ km\,s$^{-1}$ and $v_{\rm LSR}=-116$ km\,s$^{-1}$ in the
lines of Si\,{\sc ii}, C\,{\sc ii}, Al\,{\sc ii}, O\,{\sc i},
Si\,{\sc iii}, and many other ions (see Fig.\,B.3; Table B.2; also
\cite{2001Richter}, their Fig.\,2). Detailed studies of
HVC Complex C towards PG\,1259$+$593 are presented in
\cite{2001Richter}, \cite{2003Sembach}, \cite{2003Collins}, and \cite{2004Fox}.

{\it\bf Mrk\,205.}
The sightline towards the Seyfert 1 galaxy Mrk\,205 passes
high-velocity gas at negative velocities at $l=125.5$, $b=+41.7$,
associated with HVC Complex C. The available E140M data
have a very good S/N of $\sim 22$ per $3.2$ km\,s$^{-1}$-wide pixel element
at 1300 \AA. Three absorption components at $v_{\rm LSR}=-106,
-138$, and $-197$ km\,s$^{-1}$ are identified in the lines of
Si\,{\sc ii}, C\,{\sc ii} and O\,{\sc i} (Fig.\,B.1; Table B.4).
For a more detailed study of this sightline see \cite{2007Collins}.

{\it\bf 3C\,249.1.}
The sightline towards the Seyfert 1 galaxy 3C\,249.1 ($l=130.4$, $b=+38.5$) 
passes the outskirts of Complex C. With a S/N of $\sim 8$ per $3.2$ 
km\,s$^{-1}$-wide pixel element at 1260 \AA\, the quality of the 
available E140M data is relatively low. We identify one HVC 
absorption component at $v_{\rm LSR}=-135$ km\,s$^{-1}$ in the 
lines of Si\,{\sc ii}, C\,{\sc ii}, and Si\,{\sc iii} 
(see Fig.\,B.1; Table B.1). HVC absorption along this 
sightline is mentioned by \cite{2009Collins}.

{\it\bf PG\,0117$+$21.}
For the line of sight towards the quasar PG\,0117$+$21 
($l=131.8$, $b=-40.8$) only E230M data are available. We identify 
one weak HVC absorber at $v_{\rm LSR}=-134$ km\,s$^{-1}$ in the 
lines of Mg\,{\sc ii} and Fe\,{\sc ii} (see Fig.\,B.1; Table B.3). 
The S/N of the data is low ($\sim 6$ per $4.8$ km\,s$^{-1}$-\,wide pixel 
element at 2796 \AA). The HVC towards PG\,0117$+$21 apparently is not connected
with any large HVC complexes. We are not aware of any previous study that discusses
the PG\,0117$+$21 sightline with respect to HVC absorption.

{\it\bf NGC\,3516.}
The line of sight toward NGC\,3516 ($l=133.2$, $b=+37.6$) passes
through the outskirts of HVC Complex C. For this sightline 
E140M and E230M data are available, but only the E230M
data have sufficient quality to measure HVC absorption features 
(S/N$\sim 7$ per $4.8$ km\,s$^{-1}$-wide pixel element at 2800 \AA.)
One single absorption component is detetected at typical
Complex C velocities of $v_{\rm LSR}=-158$ km\,s$^{-1}$
in the lines of Mg\,{\sc ii} and Fe\,{\sc ii}
(see Fig.\,B.2; Table B.3). 
HVC absorption towards NGC\,3516 is mentioned
in the papers from \cite{2009Collins} and
\cite{2009Shull}.

{\it\bf NGC\,4151.}
Along the sightline towards NGC\,4151 ($l=155.1$, $b=+75.1$)
weak HVC absorption is observed at high positive radial
velocities in the lines of C\,{\sc ii}, Si\,{\sc ii}, Fe\,{\sc ii},
Mg\,{\sc ii}, Si\,{\sc iii}, and C\,{\sc iv} in a
single absorption component centred
at $v_{\rm LSR}=+143$ km\,s$^{-1}$ (Fig.\,B.2; Table B.3).
Both the available E140M and E230M data have good
S/N ($\sim 19$ per $3.2$ km\,s$^{-1}$-wide pixel element
at 1300 \AA). This HVC appears to be isolated without being
connected to any known large HVC complex. A detailed
study of this HVC is presented in \cite{2009Richter}.

{\it\bf Mrk\,132.}
The line of sight towards Mrk\,132 is located at $l=158.9$, $b=+48.6$, 
located between the two HVC Complexes B and C. Only E230M data are 
available to us. We find high-velocity absorption at negative velocities 
($v_{\rm LSR}=-139$ km\,s$^{-1}$ and $v_{\rm LSR}=-98$ km\,s$^{-1}$), 
as well as absorption at positive velocities ($v_{\rm LSR} \approx 80$ km\,s$^{-1}$) 
in the lines of Mg\,{\sc ii} and Fe\,{\sc ii} (see Fig.\,B.1; Table B.4). 
The best fit for the positive-velocity absorption yields two components 
blending each other. With a S/N of $\sim 10$ per $4.8$ km\,s$^{-1}$-\,wide 
pixel element at 2796 \AA \,the data are of medium quality. 
We are not away of any previous paper that discusses the properties 
of the HVCs towards Mrk\,132. 

{\it\bf PG\,0953$+$415.}
The line of sight towards the Seifert\,1
galaxy PG\,0953$+$415 is located at $l=179.8$ and $b=+51.7$ and
passes high-velocity halo gas that possibly is related
to HVC Complex M. For this sightline only E140 M data
with intermediate S/N ($\sim 9$ per
$3.2$ km\,s$^{-1}$-wide pixel element at 1300 \AA)
are available. We identify one HVC component at
$v_{\rm LSR}=-147$ km\,s$^{-1}$ in the lines of
C\,{\sc ii}, Al\,{\sc ii}, Si\,{\sc ii}, and Si\,{\sc iii}
(see Fig.\,B.2; Table B.1). No previous study exists that
analyses in detail the low-ion absorption near
$v_{\rm LSR}=-147$ km\,s$^{-1}$, but the sightline
has been studied by several authors to investigate
highly-ionized gas in the Milky Way halo at
{\it positive} radial velocities
(\cite{2005Collins}, \cite{2009Collins}; \cite{2009Fox};
\cite{2009Shull}; \cite{2000Fabian}).

{\it\bf PG\,1116$+$215.}
High-velocity absorption towards PG\,1116$+$215
($l=223.4,b=+68.2$) is seen at positive LSR velocities
between $+170$ and $+220$ km\,s$^{-1}$ in the lines
of Si\,{\sc ii}, C\,{\sc ii}, O\,{\sc i}, Fe\,{\sc ii},
Mg\,{\sc ii}, Si\,{\sc iii}, C\,{\sc iv}, and Si\,{\sc iv}
(see Fig.\,B.2; Table B.1).
This multi-phase absorber is relatively isolated, but
possibly is associated with the Magellanic Stream. For this
sightline both E140M and E230M data with relatively
good S/N ($\sim 10$ per $3.2$ km\,s$^{-1}$-wide pixel element
at 1300 \AA) are available. Two absorption components at
$v_{\rm LSR}=+184$ and $+203$ km\,s$^{-1}$ can be fitted
to the data. The HVC absorption towards PG\,1116$+$215
was studied in detail by \cite{2005Ganguly} and
\cite{2009Richter}. Other studies that discuss
the HVC absorption along this sightlines are
\cite{2004Sembach2}, \cite{2005Collins}, \cite{2009Collins}
and \cite{2006Fox}.

{\it\bf Ton\,S210.}
Along the line of sight towards Ton\,S210
($l=225.0$, $b=-83.2$) high-velocity absorption
is seen at high negative velocities between
$-140$ and $-260$ km\,s$^{-1}$. Absorption
near $-170$ km\,s$^{-1}$ is related to the
compact high-velocity cloud CHVC\,$224.0-83.4$
(\cite{2002Putman}), while the absorption
at higher velocities is of unknown origin.
Both E140M and E230M data are available, but the
S/N is low (S/N$\sim 4 $ per $3.2$ km\,s$^{-1}$-wide pixel
element at 1500 \AA). Three individual
absorption components at $v_{\rm LSR}=-172,
-207$, and $-241$ km\,s$^{-1}$ are identified
in the lines of C\,{\sc ii}, O\,{\sc i}, Si\,{\sc ii},
Mg\,{\sc ii}, Fe\,{\sc ii}, Si\,{\sc iii}, 
Si\,{\sc iv} and C\,{\sc iv} (see Fig.\,B.4; 
Table B.5). A detailed
analysis of the high-velocity gas towards
Ton\,S210 is presented in \cite{2002Sembach}
and \cite{2009Richter}.

{\it\bf HE\,0515$-$4414.}
The sightline towards HE\,0515$-$4414 is located at $l=249.6$, $b=-35.0$
and passes through a region with some scattered H\,{\sc i} clouds that
possibly are connected to the Magellanic Stream. Only E230M data
are available for this sightline, and the data have a low S/N of
$\sim 5$ per $4.8$ km\,s$^{-1}$-wide pixel element
at 2400 \AA. We identify one HVC absorption component at
$v_{\rm LSR}=+103$ km\,s$^{-1}$ in the lines of Mg\,{\sc ii} and
Fe\,{\sc ii}. Additional HVC absorption components possibly
are present near $\sim +200$ km\,s$^{-1}$, but are blended
with intergalactic absorption features (Fig.\,B.1; Table B.4).
We did not find any previous paper that discusses the HVC
absorption towards HE\,0515$-$4414.

{\it\bf PG\,1211$+$143.}
Relatively weak weak absorption at positive LSR velocities
near $v_{\rm LSR}=+180$ km\,s$^{-1}$ is seen along the line of sight
towards the Seyfert 1 galaxy PG\,1211$+$143 ($l=267.6$,
$b=+74.3$) in the lines of Si\,{\sc ii}, C\,{\sc ii},
O\,{\sc i}, Si\,{\sc iii}, and C\,{\sc iv}
(see Fig.\,B.3; Table B.3). For this
line of sight, only E140M data are available. The
S/N in the data is good ($\sim 15$ per $3.2$ km\,s$^{-1}$-wide
pixel element at 1300 \AA). Two distinct
absorption components are observed at $v_{\rm LSR}=+170$ km\,s$^{-1}$
and $+188$ km\,s$^{-1}$; the latter component is seen
predominantly in the intermediate and high ions. This
HVC is isolated, but possibly is associated with a
compact high-velocity cloud (CHVC) $\sim 1.5$ deg away
(\cite{2002deHeij}). A detailed study of this absorber
is presented in \cite{2009Richter}; other studies that
mention this HVC are \cite{2006Fox} and \cite{2009Collins}.

{\it\bf PG\,1216$+$069.}
The Seyfert 1 galaxy PG\,1216$+$069 is located at $l=281.1$, $b=+68.1$.
We find HVC absorption in the lines of Si\,{\sc ii},
Si\,{\sc iii}, and C\,{\sc iv} in three individual components at high LSR-velocities 
of $v_{\rm LSR}=+210\,...\,+270$ km\,s$^{-1}$ 
in the available E140M data (see Fig.\,B.1; Table B.3). The detected absorbers
seem to be unrelated to any known large HVC complex.
The data quality is relatively low (S/N$\sim 8$ per $3.2$ km\,s$^{-1}$-\,wide 
pixel element at 1260 \AA). 
HVC absorption towards PG\,1216$+$069 is also discussed
in \cite{2009Shull}.

{\it\bf NGC\,3783.}
The line of sight towards the Seyfert 1 galaxy
NGC\,3783 at $l=287.5$, $b=+23.0$ is a well-studied
HVC sightline that passes through the Leading Arm (LA)
of the Magellanic Stream with strong absorption
features at high positive velocities. For this
sightline both E140M and E230M data are available.
The data have superb quality with a S/N
of $\sim 25$ per $3.2$ km\,s$^{-1}$-wide pixel
element at 1300 \AA. Very strong disc and halo
absorption from local disc and halo gas
is seen in a single absorption trough in
many lines in the velocity range between $-60$
and $+120$ km\,s$^{-1}$. This absorption
component is not further considered in this
paper. Absorption from the LA is seen in
two strong components at $v_{\rm LSR}=+180$ and
$+234$ km\,s$^{-1}$ in the lines of Si\,{\sc ii},
O\,{\sc i}, Mg\,{\sc ii}, Fe\,{\sc ii}, Al\,{\sc ii},
C\,{\sc ii}, and Si\,{\sc iii} (see Fig.\,B.2;
Table B.3). There are several detailed studies
of the LA absorption towards NGC\,3783
(e.g., \cite{1998Lu}; \cite{2001Sembach}).

{\it\bf RXJ\,1230.8$+$0115.}
In the direction of RXJ\,1230.8$+$0115 ($l=291.3$, $b=+63.7$)
weak HVC absorption in the lines of Si\,{\sc ii}, C\,{\sc ii},
O\,{\sc i} and Si\,{\sc iii} is observed in two individual
clouds at high positive velocities near $+100$ and $+300$
km\,s$^{-1}$ (only E140M data are available; Fig.\,B.27; Table
B.3). The data are of intermediate quality with a S/N
of $\sim 7$ per $3.2$ km\,s$^{-1}$-wide pixel element
at 1300 \AA.
For each of the two clouds, the HVC absorption can be fitted
with a single absorption component centred at $v_{\rm LSR}=+111$ and
$+295$ km\,s$^{-1}$, respectively (for C\,{\sc ii}
we add another component at $v_{\rm LSR}=+292$ km\,s$^{-1}$).
Both HVCs do not appear to be
associated with any prominent HVC complex, but obviously respresent
isolated gaseous halo strucures. A recent study of this HVC
sightline is presented by \cite{2009Richter}.

{\it\bf PKS\,0312$-$770.}
The line of sight towards the Seyfert\,1 galaxy PKS\,0312$-$770
($l=293.4$, $b=-37.6$) is known to pass the so-called
``Magellanic Bridge'' (MB), an extended gaseous structure
that connects the Large Magellanic Cloud
(LMC) and the Small Magellanic Cloud (SMC; see
\cite{1963Hindman}). The MB is believed to be
locted at a distance of $\sim 50-60$ kpc
(\cite{2003Harries}). Thus, the MB does not represent
a ``classical'' Galactic HVC but rather is a tidal
feature and gas component of the Magellanic system.
For this sightline both E140M and E230M spectra are available,
but the average S/N is relatively low (S/N$\sim 6$ per
$3.2$ km\,s$^{-1}$-wide pixel element at 1500 \AA).
Strong high-velocity absorption from gas in the MB
is visible in the lines of C\,{\sc ii}, O\,{\sc i}, Si\,{\sc ii},
Si\,{\sc iii}, Fe\,{\sc ii}, and Mg\,{\sc ii}. The STIS data
indicate two main absorption components centred at
$v_{\rm LSR}=+174$ km\,s$^{-1}$ and v$_{\rm LSR}=+224$ km\,s$^{-1}$
(see Fig.\,B.3; Table B.1). A detailed analysis of this
sightline and the MB absorption is presented in
\cite{2009Misawa}.

%

\newpage

\section{Supplementrary tables and figures}


\begin{table}[ht]
\small
\caption[]{Summary of HVC absorption-line measurements$^{\rm a}$ - part I}
\vspace{0.5cm}
\label{data1}
\begin{tabular}{l r l l r } \hline\hline
Sightline & $v_{\rm LSR}$ && log $N$ & $b$ \\
&(km s$^{-1}$)&Ion&($N$ in cm$^{-2}$)&(km s$^{-1}$) \\\hline
PKS\,0312$-$770&$+174$&Si\,{\sc ii}&14.92$\pm$0.95&14.2$\pm$5.9\\
        &&C\,{\sc ii}&13.82$\pm$1.06&14.2$\pm$5.9\\
        &&O\,{\sc i}&14.99$\pm$0.14&14.2$\pm$5.9\\
        &&Fe\,{\sc ii}&13.97$\pm$0.26&14.2$\pm$5.9\\
        &&Mg\,{\sc ii}&15.22$\pm$0.34&14.2$\pm$5.9\\
        &&Si\,{\sc iii}& ... & ... \\
        &$+224$&Si\,{\sc ii}&14.73$\pm$0.33&16.6$\pm$3.1\\
        &&O\,{\sc i}&14.76$\pm$1.02&16.6$\pm$3.1\\
        &&Mg\,{\sc ii}&15.08$\pm$0.29&16.6$\pm$3.1\\
        &&Fe\,{\sc ii}&15.05$\pm$0.41&16.6$\pm$3.1\\
        &&Si\,{\sc iii}& ... & ... \\
PG\,0953$+$415&$-147$&Si\,{\sc ii}&12.60$\pm$0.06&8.5$\pm$1.5 \\
        &&C\,{\sc ii}&13.52$\pm$0.07&8.5$\pm$1.5 \\
        &&Si\,{\sc iii}&12.77$\pm$0.09&11.9$\pm$2.5 \\
3c249.1 &$-135$&Si\,{\sc ii}&12.37$\pm$0.08&6.6$\pm$1.7\\
        &&C\,{\sc ii}&13.75$\pm$0.07&6.6$\pm$1.7\\
        &&Si\,{\sc iii}&12.59$\pm$0.18&8.6$\pm$5.4\\
3c351&$-193$&Si\,{\sc ii}&13.20$\pm$0.06&7.9$\pm$1.3\\
        &&C\,{\sc ii}& ... & ... \\
        &&O\,{\sc i}&13.89$\pm$0.05&7.9$\pm$1.3\\
        &&Al\,{\sc ii}&12.17$\pm$0.07&7.9$\pm$1.3\\
        &&Si\,{\sc iii}&14.52$\pm$0.59 &8.6$\pm$1.2\\
        &&Si\,{\sc iv}&12.62$\pm$0.10&9.5$\pm$3.7\\
        &&C\,{\sc iv}&13.23$\pm$0.08&9.5$\pm$3.7\\
        &$-130$&Si\,{\sc ii}&13.80$\pm$0.05&9.4$\pm$0.9\\
        &&C\,{\sc ii}&14.31$\pm$0.26 &9.4$\pm$0.9\\
        &&O\,{\sc i}&14.58$\pm$0.07&9.4$\pm$0.9\\
        &&Al\,{\sc ii}&12.64$\pm$0.06&9.4$\pm$0.9\\
        &$-101$&Si\,{\sc ii}&14.19$\pm$0.06&11.6$\pm$1.0\\
        &&O\,{\sc i}&14.30$\pm$0.07&11.6$\pm$1.0\\
        &&Al\,{\sc ii}&12.49$\pm$0.07&11.6$\pm$1.0\\
	&$-82$&Si\,{\sc ii}&13.51$\pm$0.10&7.8$\pm$1.3\\
	&&Al\,{\sc ii}&12.54$\pm$0.12&7.8$\pm$1.3\\
	&&O\,{\sc i}&15.14$\pm$0.26&7.8$\pm$1.3\\
	&&Si\,{\sc iv}&12.79$\pm$0.07&17.8$\pm$0.8\\
        &&C\,{\sc iv}&13.60$\pm$0.06&17.8$\pm$0.8\\
	&$-76$&Al\,{\sc ii}&12.73$\pm$0.41&4.0$\pm$1.3\\
PG\,1116$+$215&$+184$&Si\,{\sc ii}&13.80$\pm$0.03&11.3$\pm$0.7\\
        &&C\,{\sc ii}&15.29$\pm$0.09&11.3$\pm$0.7\\
        &&O\,{\sc i}&13.91$\pm$0.05&11.3$\pm$0.7\\
        &&Fe\,{\sc ii}&13.36$\pm$0.19&11.3$\pm$0.7\\
        &&Mg\,{\sc ii}&12.99$\pm$0.24&11.3$\pm$0.7\\
        &&Si\,{\sc iii}&13.23$\pm$0.25&20.0$\pm$4.0\\
        &&Si\,{\sc iv}&12.99$\pm$0.05&14.6$\pm$2.3\\
        &&C\,{\sc iv}&13.76$\pm$0.07&14.6$\pm$2.3\\
        &$+203$&Si\,{\sc ii}&12.64$\pm$0.27&7.9$\pm$3.3\\
        &&C\,{\sc ii}&13.76$\pm$0.05&7.9$\pm$3.3\\
        &&O\,{\sc i}&13.66$\pm$0.07&7.9$\pm$3.3\\
        &&Fe\,{\sc ii}&12.57$\pm$0.48&7.9$\pm$3.3\\
        &&Mg\,{\sc ii}&12.74$\pm$0.41&7.9$\pm$3.3\\
        &&Si\,{\sc iii}&14.42$\pm$1.13&4.0$\pm$3.6\\
\hline
\end{tabular}
\noindent
\\
{$^{\rm a}$ We do not list column-density limits based on saturated lines.}
\end{table}


\begin{table}[ht]
\small
\caption[]{Summary of HVC absorption-line measurements$^{\rm a}$ - part II}
\vspace{0.5cm}
\begin{tabular}{l r l l r } \hline\hline
Sightline & $v_{\rm LSR}$ && log $N$ & $b$ \\
&(km s$^{-1}$)&Ion&($N$ in cm$^{-2}$)&(km s$^{-1}$) \\\hline
PG\,1259$+$593&$-134$&Si\,{\sc ii}&14.60$\pm$0.56&5.6$\pm$1.6\\
        &&C\,{\sc ii}&15.20$\pm$0.51&5.6$\pm$1.6\\
        &&Al\,{\sc ii}&13.44$\pm$0.44&5.6$\pm$1.6\\
        &&O\,{\sc i}&14.31$\pm$0.52&5.6$\pm$1.6\\
        &$-116$&Si\,{\sc ii}&13.72$\pm$0.16&10.8$\pm$3.7\\
        &&O\,{\sc i}&14.69$\pm$0.32&10.8$\pm$3.7\\
        &&Al\,{\sc ii}&12.58$\pm$0.12&10.8$\pm$3.7\\
PG\,1634$+$706&$-195$&Fe\,{\sc ii}&13.19$\pm$0.05&41.1$\pm$2.3\\
        &&Mg\,{\sc ii}&13.30$\pm$0.03&41.1$\pm$2.3\\
        &$-164$&Fe\,{\sc ii}&13.15$\pm$0.07&8.2$\pm$1.7\\
        &&Mg\,{\sc ii}&13.08$\pm$0.12&8.2$\pm$1.7\\
        &$-125$&Fe\,{\sc ii}&13.36$\pm$0.03&22.0$\pm$1.3\\
        &&Mg\,{\sc ii}&13.28$\pm$0.02&22.0$\pm$1.3\\
PG\,1718$+$481 &$-197$&Fe\,{\sc ii} &12.80$\pm$0.05 &10.2$\pm$2.5\\
        &$-128$&Fe\,{\sc ii} &13.17$\pm$0.04 &25.4$\pm$3.7\\
        &$-81$&Fe\,{\sc ii} &13.37$\pm$0.05 &9.7$\pm$1.0\\
PHL\,1811&$-351$&Si\,{\sc iii}& ... & ... \\
	&&Si\,{\sc iv}&12.68$\pm$0.13&7.3$\pm$3.6\\
	&&C\,{\sc iv}&14.57$\pm$0.19&7.3$\pm$3.6\\
	&$-263$&C\,{\sc ii}&13.52$\pm$0.16&11.4$\pm$5.6\\
        &&Si\,{\sc iii}&13.14$\pm$0.07&22.9$\pm$3.3\\
        &$-240$&C\,{\sc ii}&13.13$\pm$0.37&6.9$\pm$4.3\\
        &&C\,{\sc iv}&13.43$\pm$0.11&6.9$\pm$4.3\\
        &$-206$&Si\,{\sc ii}&13.48$\pm$0.06&13.7$\pm$2.0\\
        &&C\,{\sc ii}&14.25$\pm$0.10&13.7$\pm$2.0\\
        &&O\,{\sc i}&14.15$\pm$0.05&13.7$\pm$2.0\\
        &&Al\,{\sc ii}&12.63$\pm$0.13&13.7$\pm$2.0\\
        &&C\,{\sc iv}&13.41$\pm$0.08&13.7$\pm$2.0\\
        &&Si\,{\sc iv}&12.92$\pm$0.09&13.7$\pm$2.0\\
        &&Si\,{\sc iii}&13.61$\pm$0.21&13.7$\pm$2.0\\
        &$-163$&Si\,{\sc ii}&12.70$\pm$0.09&8.7$\pm$1.6\\
        &&C\,{\sc ii}&13.92$\pm$0.10&8.7$\pm$1.6\\
        &&Al\,{\sc ii}&12.16$\pm$0.19&8.7$\pm$1.6\\
        &&Si\,{\sc iii}&13.39$\pm$0.23&13.5$\pm$3.5\\
        &&Si\,{\sc iv}&13.48$\pm$0.27&8.3$\pm$2.3\\
        &&C\,{\sc iv}&15.00$\pm$0.17&8.3$\pm$2.3\\
H\,1821$+$643&$-146$&Si\,{\sc iv}&12.40$\pm$0.13&3.8$\pm$1.6\\
        &$-126$&Si\,{\sc ii}&14.21$\pm$0.02&18.5$\pm$0.9 \\
        &&C\,{\sc ii}&14.62$\pm$0.03&18.5$\pm$0.9 \\
        &&O\,{\sc i}&14.78$\pm$0.04&18.5$\pm$0.9 \\
        &&Al\,{\sc ii}&13.09$\pm$0.04&18.5$\pm$0.9 \\
        &&Si\,{\sc iii}& ... & ... \\
        &&C\,{\sc iv}&13.47$\pm$0.06&14.2$\pm$4.8 \\
        &&Si\,{\sc iv}&13.03$\pm$0.06&14.2$\pm$4.8 \\
        &$-84$&Si\,{\sc ii}&14.28$\pm$0.08&9.4$\pm$1.0 \\
        &&C\,{\sc ii}& ... & ... \\
        &&O\,{\sc i}&15.33$\pm$0.25&9.4$\pm$1.0 \\
        &&Al\,{\sc ii}&13.24$\pm$0.20&9.4$\pm$1.0 \\
\hline
\end{tabular}
\label{data2}
\noindent
\\
{$^{\rm a}$ We do not list column-density limits based on saturated lines.}
\end{table}


\begin{table}[ht]
\small
\caption[]{Summary of HVC absorption-line measurements$^{\rm a}$ - part III}
\vspace{0.5cm}
\begin{tabular}{l r l l r } \hline\hline
Sightline & $v_{\rm LSR}$ && log $N$ & $b$ \\
&(km s$^{-1}$)&Ion&($N$ in cm$^{-2}$)&(km s$^{-1}$) \\\hline
PG\,0117$+$21&$-134$&Mg\,{\sc ii}&12.72$\pm$0.06&11.1$\pm$2.1\\
        &&Fe\,{\sc ii}&12.73$\pm$0.10&11.1$\pm$2.1\\
PG\,1211$+$143&$+170$&Si\,{\sc ii}&12.64$\pm$0.04&6.7$\pm$0.8\\
        &&C\,{\sc ii}&13.61$\pm$0.03&6.7$\pm$0.8\\
        &&O\,{\sc i}&13.14$\pm$0.06&6.7$\pm$0.8\\
        &&Si\,{\sc iii}&12.28$\pm$0.14&9.2$\pm$3.5\\
        &$+188$&C\,{\sc iv}&12.81$\pm$0.13&14.3$\pm$6.0\\
        &&Si\,{\sc iii}&12.11$\pm$0.19&6.5$\pm$3.2\\
        &&C\,{\sc ii}&13.24$\pm$0.07&11.6$\pm$2.1\\
PG\,1216$+$069&$+267$&Si\,{\sc ii}&12.98$\pm$0.08&12.7$\pm$2.6\\
        &$+256$&Si\,{\sc ii}&12.60$\pm$0.17&4.8$\pm$2.5\\
        &&Si\,{\sc iii}&13.16$\pm$0.09&19.2$\pm$2.9\\
        &&C\,{\sc iv}&13.65$\pm$0.11&19.1$\pm$6.1\\
        &$+212$&Si\,{\sc ii}&12.60$\pm$0.09&11.3$\pm$3.1\\
        &&Si\,{\sc iii}&12.69$\pm$0.13&8.1$\pm$2.5\\
        &&C\,{\sc iv}&13.23$\pm$0.14&8.1$\pm$2.5\\
PG\,1444$+$407&$-88$&C\,{\sc iv}&13.82$\pm$0.12&24.3$\pm$8.3\\
	&$-81$&Si\,{\sc ii}&13.69$\pm$0.25&7.5$\pm$2.4\\
        &&C\,{\sc ii}&... &...\\
        &&O\,{\sc i}&14.16$\pm$0.05&7.5$\pm$2.4\\
PG\,1630$+$377&$-155$&Fe\,{\sc ii}&12.60$\pm$0.15&7.5$\pm$2.3\\
        &&Mg\,{\sc ii}&12.47$\pm$0.09&7.5$\pm$2.3\\
        &$-99$&Fe\,{\sc ii}& ... & ... \\
        &&Mg\,{\sc ii}& ... & ... \\
        &$-64$&Fe\,{\sc ii}&13.49$\pm$0.12&11.5$\pm$9.8\\
        &&Mg\,{\sc ii}&13.72$\pm$0.45&11.5$\pm$9.8\\
RXJ\,1230.8$+$0115&$+111$& Si\,{\sc ii}&12.66$\pm$0.07&12.6$\pm$2.9\\
        &&C\,{\sc ii}&13.74$\pm$0.04&12.6$\pm$2.9\\
        &&Si\,{\sc iii}&12.77$\pm$0.10&12.6$\pm$2.9\\
        &$+292$&C\,{\sc ii}&13.60$\pm$0.10&18.6$\pm$5.0\\
        &$+295$&Si\,{\sc ii}&13.55$\pm$0.13&3.7$\pm$0.5\\
        &&C\,{\sc ii}&14.34$\pm$0.36&3.7$\pm$0.5\\
        &&O\,{\sc i}&13.77$\pm$0.12&3.7$\pm$0.5\\
        &&Si\,{\sc iii}&12.66$\pm$0.11&13.0$\pm$4.2\\
NGC\,3516&$-158$&Mg\,{\sc ii}&12.94$\pm$0.07&27.6$\pm$5.7\\
        &&Fe\,{\sc ii}&13.25$\pm$0.07&27.6$\pm$5.7\\
NGC\,3783&$+180$&Si\,{\sc ii}&13.31$\pm$0.03&11.2$\pm$0.6\\
        &&O\,{\sc i}&14.84$\pm$0.04&11.2$\pm$0.6\\
        &&Mg\,{\sc ii}&12.88$\pm$0.05&11.2$\pm$0.6\\
        &&Fe\,{\sc ii}&12.86$\pm$0.08&11.2$\pm$0.6\\
        &&C\,{\sc ii}&14.57$\pm$0.04&11.2$\pm$0.6\\
        &&Si\,{\sc iii}&12.60$\pm$0.04&15.2$\pm$2.0\\
        &$+234$&Si\,{\sc ii}&13.88$\pm$0.04&17.4$\pm$0.9\\
        &&C\,{\sc ii}& ... & ... \\
        &&Mg\,{\sc ii}&13.61$\pm$0.05&17.4$\pm$0.9\\
        &&Al\,{\sc ii}&12.72$\pm$0.04&17.4$\pm$0.9\\
        &&Fe\,{\sc ii}&13.84$\pm$0.06&17.4$\pm$0.9\\
        &&O\,{\sc i}&15.13$\pm$0.03&17.4$\pm$0.9\\
        &&Si\,{\sc iii}&12.81$\pm$0.03&19.0$\pm$1.8\\
NGC\,4151&$+143$&Si\,{\sc ii}&12.53$\pm$0.04&6.4$\pm$1.0\\
        &&C\,{\sc ii}&13.56$\pm$0.04&6.4$\pm$1.0\\
        &&Mg\,{\sc ii}&12.23$\pm$0.04&6.4$\pm$1.0\\
        &&Fe\,{\sc ii}&12.00$\pm$0.04&6.4$\pm$1.0\\
        &&C\,{\sc iv}&13.56$\pm$0.06&16.2$\pm$2.5\\
        &&Si\,{\sc iii}&12.71$\pm$0.04&11.3$\pm$0.8\\
        &&Fe\,{\sc ii}&12.38$\pm$0.28&13.2$\pm$8.1\\
\hline
\end{tabular}
\label{data3}
\noindent
\\
{$^{\rm a}$ We do not list column-density limits based on saturated lines.}
\end{table}


\begin{table}[ht]
\small
\caption[]{Summary of HVC absorption-line measurements$^{\rm a}$ - part IV}
\vspace{0.5cm}
\begin{tabular}{l r l l r } \hline\hline
Sightline & $v_{\rm LSR}$ && log $N$ & $b$ \\
&(km s$^{-1}$)&Ion&($N$ in cm$^{-2}$)&(km s$^{-1}$) \\\hline
NGC\,7469&$-366$&Si\,{\sc ii}&13.07$\pm$0.05&17.0$\pm$1.8\\
        &&C\,{\sc ii}&13.91$\pm$0.04&17.0$\pm$1.8\\
        &&O\,{\sc i}&13.89$\pm$0.09&17.0$\pm$1.8\\
        &$-335$&Si\,{\sc ii}&13.29$\pm$0.15&5.3$\pm$1.3\\
        &&C\,{\sc ii}&15.07$\pm$0.32&5.3$\pm$1.3\\
        &&O\,{\sc i}&14.11$\pm$0.17&5.3$\pm$1.3\\
        &&Si\,{\sc iii}& ... &...\\
        &&Si\,{\sc iv}&13.37$\pm$0.06&30.1$\pm$5.5\\
        &&C\,{\sc iv}&13.80$\pm$0.04&30.1$\pm$5.5\\
	&$-293$&Si\,{\sc iii}&... &...\\
	&&Si\,{\sc iv}&12.55$\pm$0.23&10.1$\pm$4.3\\
        &&C\,{\sc iv}&13.12$\pm$0.14&10.1$\pm$4.3\\
        &$-251$&Si\,{\sc ii}&12.88$\pm$0.05&12.3$\pm$1.7\\
	&&C\,{\sc ii}&13.25$\pm$0.08&12.3$\pm$1.7\\
	&&Si\,{\sc iii}&... &...\\
	&&Si\,{\sc iv}&12.60$\pm$0.11&6.2$\pm$2.3\\
        &&C\,{\sc iv}&13.22$\pm$0.08&6.2$\pm$2.3\\
        &$-185$&C\,{\sc ii}&12.86$\pm$0.60&1.7$\pm$4.1\\
        &&Si\,{\sc iii}& ... & ...\\
	&&C\,{\sc iv}&13.67$\pm$0.06&11.5$\pm$2.5\\
	&&Si\,{\sc iv}&12.88$\pm$0.08&11.5$\pm$2.5\\
HE\,0515$-$4414&$+103$&Fe\,{\sc ii}&12.88$\pm$0.09&11.3$\pm$0.9\\
        &&Mg\,{\sc ii}&12.88$\pm$0.03&11.3$\pm$0.9\\
PKS\,2155$-$304&$-280$&Si\,{\sc iii}& 12.12$\pm$0.12&13.1$\pm$4.6\\
        &&Si\,{\sc iv}&12.36$\pm$0.07&10.5$\pm$1.4\\
        &&C\,{\sc iv}&13.19$\pm$0.04&10.5$\pm$1.4\\
        &$-254$&C\,{\sc iv}&12.89$\pm$0.09&5.5$\pm$1.7\\
	&&Si\,{\sc iii}&11.69$\pm$0.21&6.2$\pm$3.3\\
        &$-232$&C\,{\sc iv}&12.88$\pm$0.11&10.8$\pm$3.8\\
	&&Si\,{\sc iii}&11.67$\pm$0.22&9.3$\pm$5.0\\
        &$-157$&Si\,{\sc iv}&12.21$\pm$0.12&8.4$\pm$3.4\\
        &&C\,{\sc iv}&12.91$\pm$0.26&8.4$\pm$3.4\\
        &$-135$&Si\,{\sc iv}&12.54$\pm$0.07&13.8$\pm$6.8\\
        &&C\,{\sc iv}&12.88$\pm$0.27&13.8$\pm$6.8\\
        &&Si\,{\sc ii}&12.63$\pm$0.02&9.6$\pm$0.8\\
        &&C\,{\sc ii}&13.75$\pm$0.08&9.6$\pm$0.8\\
        &&Si\,{\sc iii}&13.15$\pm$0.05&21.7$\pm$2.0\\
        &$-111$&Si\,{\sc ii}&11.79$\pm$0.15&1.0$\pm$1.6\\
        &&C\,{\sc ii}&13.17$\pm$0.15&1.0$\pm$1.6\\
Mrk\,132 &$+76$&Mg\,{\sc ii}&12.63$\pm$0.84&5.8$\pm$3.0\\
        &&Fe\,{\sc ii}&13.28$\pm$0.12&5.8$\pm$3.0\\
	&$+85$&Mg\,{\sc ii}&12.95$\pm$0.45&8.2$\pm$4.5\\
	&$-139$&Mg\,{\sc ii}&13.10$\pm$0.08&14.1$\pm$2.5\\
	&&Fe\,{\sc ii}&13.28$\pm$0.07&14.1$\pm$2.5\\
	&$-98$&Mg\,{\sc ii}&12.87$\pm$0.11&17.5$\pm$6.9\\
	&&Fe\,{\sc ii}&13.05$\pm$0.08&17.5$\pm$6.9\\
Mrk\,205&$-197$&Si\,{\sc ii}&13.41$\pm$0.03&8.8$\pm$0.7\\
        &&C\,{\sc ii}&14.59$\pm$0.08&8.8$\pm$0.7\\
        &&O\,{\sc i}&14.77$\pm$0.03&8.8$\pm$0.7\\
        &$-138$&Si\,{\sc ii}&13.13$\pm$0.04&9.3$\pm$0.9\\
        &&C\,{\sc ii}&14.14$\pm$0.06&9.3$\pm$0.9\\
        &&O\,{\sc i}&14.00$\pm$0.02&9.3$\pm$0.9\\
        &$-106$&Si\,{\sc ii}&12.81$\pm$0.11&7.4$\pm$2.1\\
        &&C\,{\sc ii}&13.90$\pm$0.07&7.4$\pm$2.1\\
        &&O\,{\sc i}&13.57$\pm$0.04&7.4$\pm$2.1\\
\hline
\end{tabular}
\label{data4}
\noindent
\\
{$^{\rm a}$ We do not list column-density limits based on saturated lines.}
\end{table}


\begin{table}[ht]
\small
\begin{center}
\caption[]{Summary of HVC absorption-line measurements$^{\rm a}$ - part V}
\vspace{0.5cm}
\begin{tabular}{l r l l r } \hline\hline
Sightline & $v_{\rm LSR}$ && log $N$ & $b$ \\
&(km s$^{-1}$)&Ion&($N$ in cm$^{-2}$)&(km s$^{-1}$) \\\hline
Mrk\, 279&$-179$&Si\,{\sc iv}&11.87$\pm$0.19&3.3$\pm$3.5\\
	&&Si\,{\sc iii}& ...  & ...\\
        &$-161$&Si\,{\sc ii}&13.76$\pm$0.09&33.6$\pm$2.5\\
        &&Al\,{\sc ii}&12.59$\pm$0.16&33.6$\pm$2.5\\
        &&C\,{\sc ii}& ... & ... \\
        &&O\,{\sc i}&14.69$\pm$0.05&33.6$\pm$2.5\\
        &&Si\,{\sc iii}& ... & ... \\
        &&Si\,{\sc iv}&12.74$\pm$0.08&33.6$\pm$2.5\\
        &$-145$&Si\,{\sc ii}&13.89$\pm$0.06&13.1$\pm$1.4\\
        &&C\,{\sc ii}&13.59$\pm$0.20&13.1$\pm$1.4\\
        &&O\,{\sc i}&14.49$\pm$0.12&13.1$\pm$1.4\\
        &&Al\,{\sc ii}&12.69$\pm$0.15&13.1$\pm$1.4\\
        &&Si\,{\sc iii}& ... & ... \\
Mrk\,290 &$-128$& Fe\,{\sc ii}& ... & ... \\
        &$-97$& Fe\,{\sc ii}& ... & ... \\
        &$-78$ & Fe\,{\sc ii}& ... & ... \\
Mrk\,509&$-311$&Si\,{\sc iii}&12.93$\pm$0.15&12.7$\pm$5.7\\
        &$-287$&Si\,{\sc iii}&14.23$\pm$0.77&4.7$\pm$4.0\\
        &&Si\,{\sc iv}&13.01$\pm$0.16&7.6$\pm$2.1\\
        &$-273$&C\,{\sc ii}&13.90$\pm$0.06&27.1$\pm$4.3\\
        &&C\,{\sc iv}&14.16$\pm$0.04&27.1$\pm$5.8\\
        &&Si\,{\sc iv}&13.06$\pm$0.13&27.1$\pm$5.8\\
        &$-263$&Si\,{\sc iii}&12.31$\pm$0.21&5.2$\pm$3.2\\
Ton\,S210&$-241$&Si\,{\sc iii}&12.83$\pm$0.16&7.5$\pm$2.1\\
        &&C\,{\sc ii}&13.06$\pm$0.10&5.6$\pm$2.1\\
	&&Si\,{\sc iv}&12.81$\pm$0.10&5.6$\pm$1.9\\
        &&C\,{\sc iv}&14.35$\pm$0.33&5.6$\pm$1.9\\
        &$-207$&C\,{\sc ii}&13.16$\pm$0.19&4.5$\pm$2.8\\
        &&Mg\,{\sc ii}&12.85$\pm$0.35&4.5$\pm$2.8\\
	&&Fe\,{\sc ii}&12.79$\pm$0.25&4.5$\pm$2.8\\
	&&Si\,{\sc iii}&12.55$\pm$0.66&9.0$\pm$5.6\\
	&&C\,{\sc iv}&13.20$\pm$0.16&9.0$\pm$5.6\\
        &$-172$&Si\,{\sc ii}&13.81$\pm$0.08&8.3$\pm$1.3\\
        &&C\,{\sc ii}& ... & ... \\
        &&O\,{\sc i}&14.05$\pm$0.08&8.3$\pm$1.3\\
	&&Mg\,{\sc ii}&12.94$\pm$0.20&8.3$\pm$1.3\\
	&&Fe\,{\sc ii}&13.62$\pm$0.24&8.3$\pm$1.3\\
        &&Si\,{\sc iii}&13.54$\pm$0.44&14.4$\pm$11.1\\
	&&C\,{\sc iv}&13.15$\pm$0.16&14.4$\pm$11.1\\
\hline
\end{tabular}
\label{data5}
\end{center}
{$^{\rm a}$ We do not list column-density limits based on saturated lines.}
\end{table}


%

\newpage
\begin{figure*}[ht!]
\begin{center}
\resizebox{0.9\hsize}{!}{\includegraphics{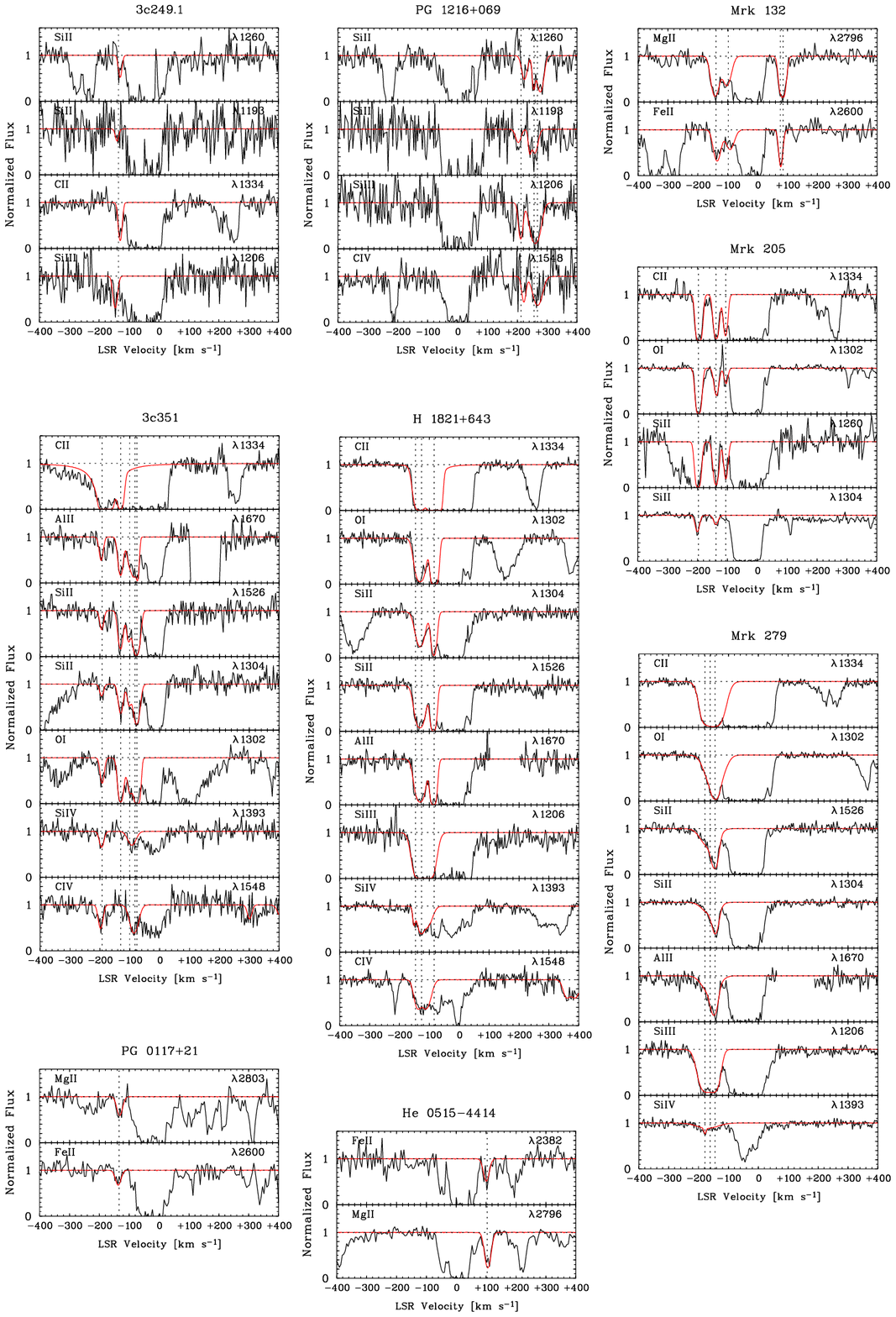}}
\end{center}
\caption[]{Continuum-normalized absorption profiles of low and high
ions towards different QSO sightlines recorded with the
E140M and E230M echelle gratings of STIS. The data are plotted
against the LSR radial velocity. Identified HVC absorption components
are marked with dashed lines. The red plotted lines render the best
Voigt-profile fit.}
\end{figure*}

%

\begin{figure*}[ht!]
\begin{center}
\resizebox{0.9\hsize}{!}{\includegraphics{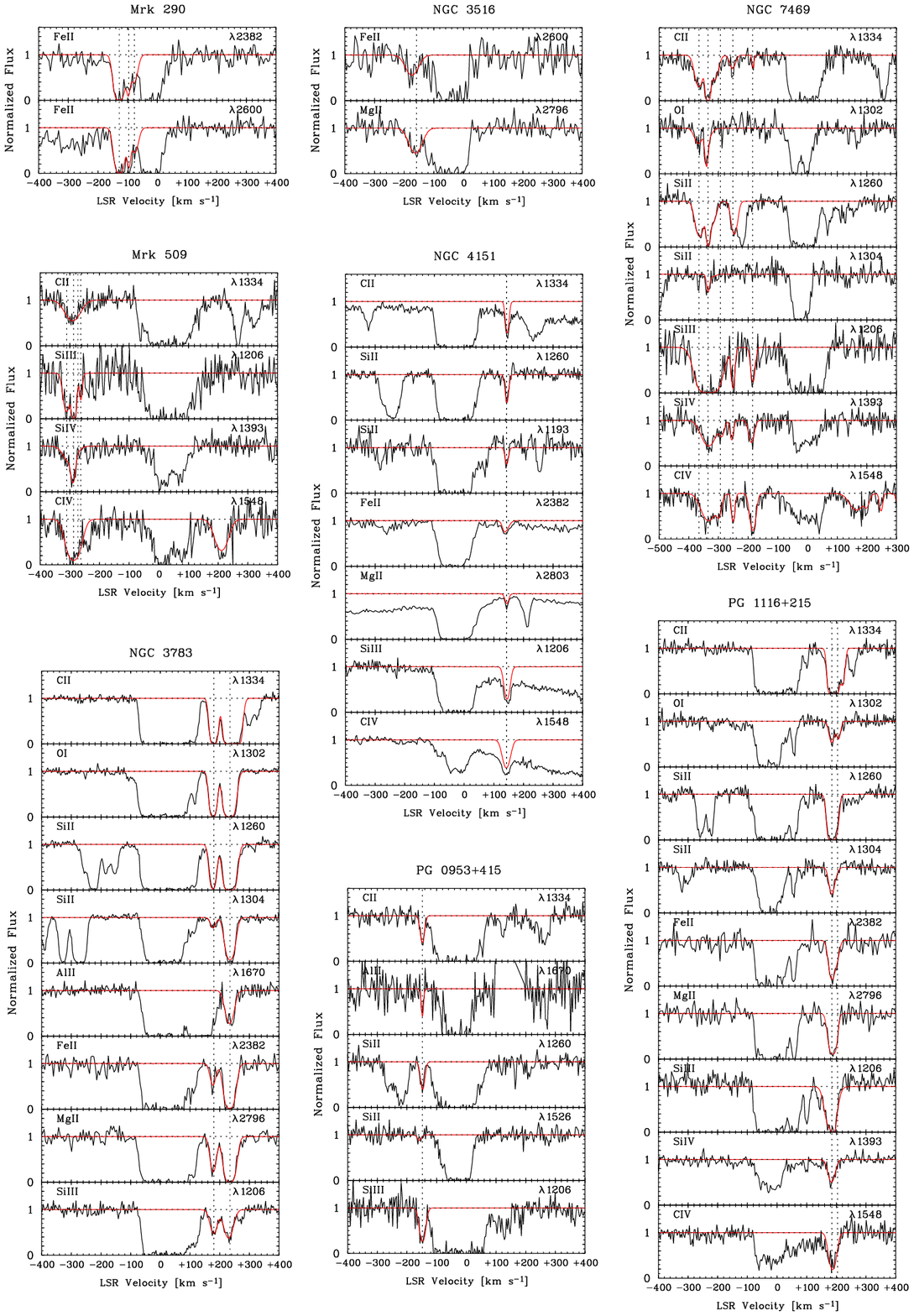}}
\end{center}
\caption[]{Additional continuum-normalized absorption profiles of low and high
ions towards different QSO sightlines recorded with the
E140M and E230M echelle gratings of STIS. The data are plotted
against the LSR radial velocity. Identified HVC absorption components
are marked with dashed lines. The red plotted lines render the best
Voigt-profile fit.}
\end{figure*}

%

\begin{figure*}[ht!]
\begin{center}
\resizebox{0.9\hsize}{!}{\includegraphics{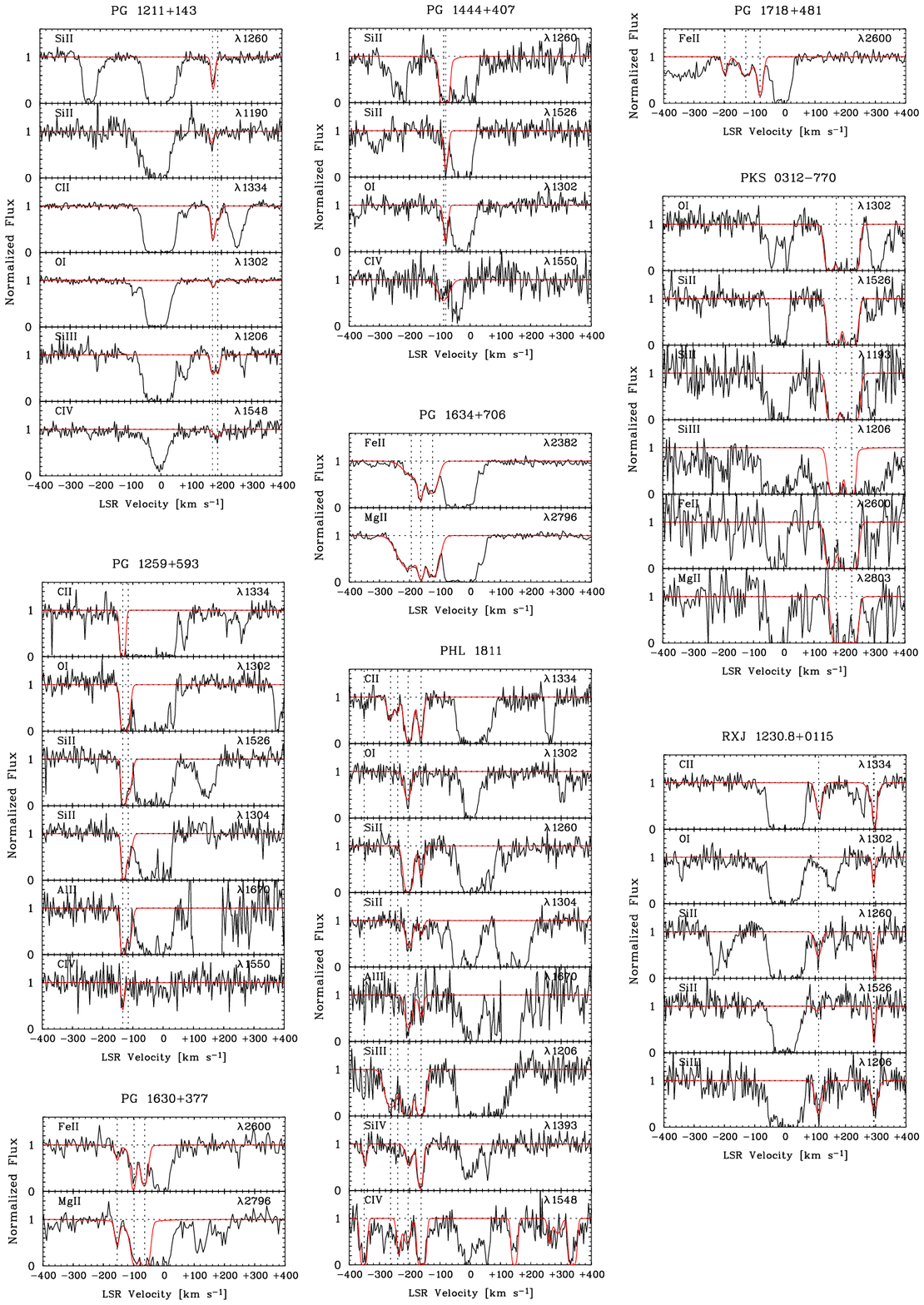}}
\end{center}
\caption[]{Additional continuum-normalized absorption profiles of low and high
ions towards different QSO sightlines recorded with the
E140M and E230M echelle gratings of STIS. The data are plotted
against the LSR radial velocity. Identified HVC absorption components
are marked with dashed lines. The red plotted lines render the best
Voigt-profile fit.}
\end{figure*}

%

\begin{figure*}[ht!]
\begin{center}
\resizebox{0.6\hsize}{!}{\includegraphics{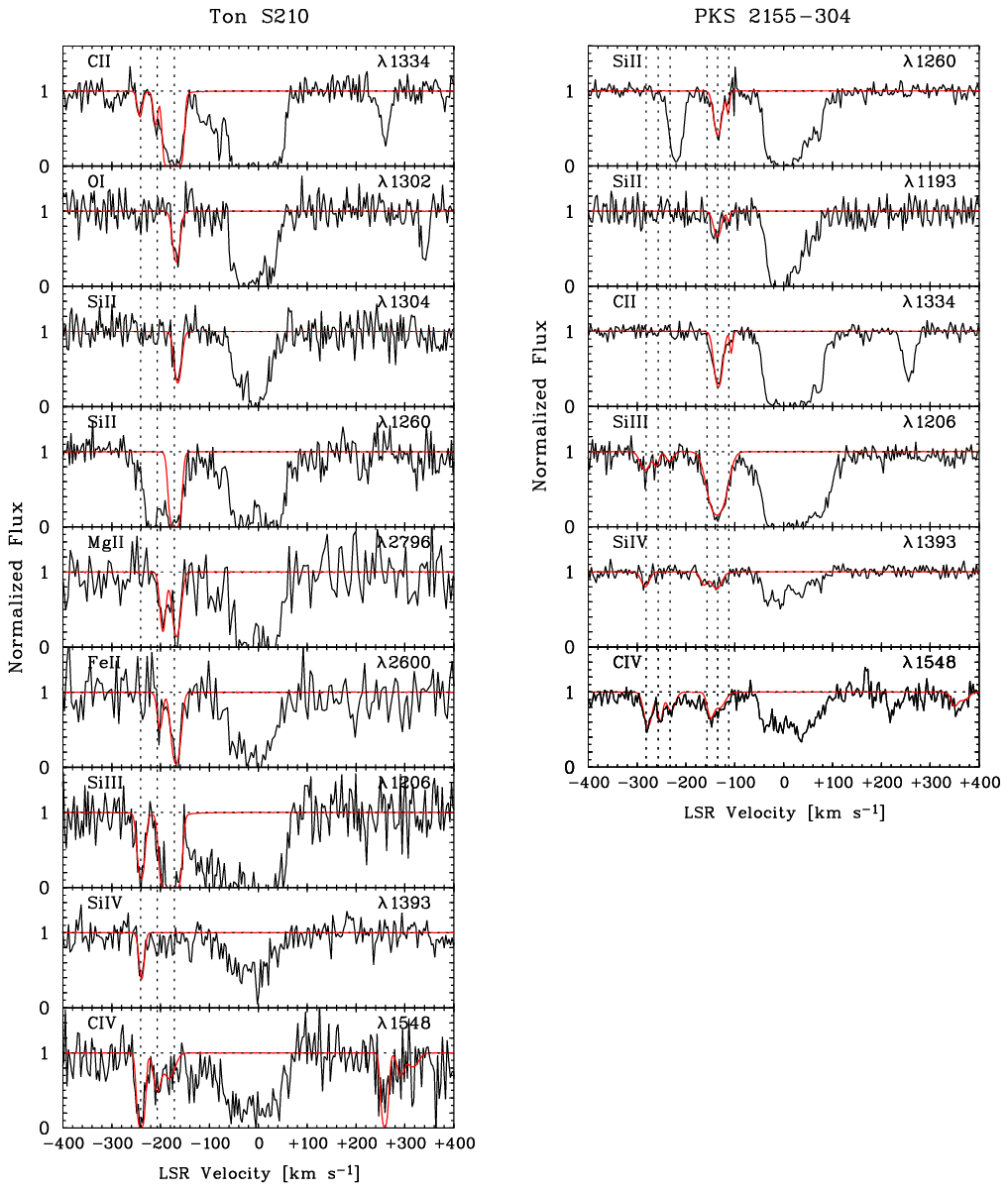}}
\end{center}
\caption[]{Additional continuum-normalized absorption profiles of low and high
ions towards different QSO sightlines recorded with the
E140M and E230M echelle gratings of STIS. The data are plotted
against the LSR radial velocity. Identified HVC absorption components
are marked with dashed lines. The red plotted lines render the best
Voigt-profile fit.}
\end{figure*}

%

\end{document}